\documentclass[iop]{emulateapj-rtx4}
\lefthead{Collins et al. 2014} \righthead{KELT-6\lowercase{b}}
\newcommand{\be}{\begin{equation}}
\newcommand{\ee}{\end{equation}}
\newcommand{\bea}{\begin{eqnarray}}
\newcommand{\eea}{\end{eqnarray}}
\newcommand{\mjup}{{\rm M}_{\rm Jup}}
\newcommand{\rjup}{{\rm R}_{\rm Jup}}
\newcommand{\teff}{\ensuremath{T_{\rm eff}}}
\newcommand{\teq}{\ensuremath{T_{\rm eq}}}
\newcommand{\feh}{{\rm [Fe/H]}}

\newcommand{\loggp}{\ensuremath{\log{g_{\rm P}}}}
\newcommand{\loggstar}{\ensuremath{\log{g_\star}}}
\newcommand{\pchisq}{\ensuremath{P\left(>\chisq \right)}}
\newcommand{\chisq}{\ensuremath{\chi^{\,2}}}
\newcommand{\vsini}{\ensuremath{\,{v\sin{i_\star}}}}
\newcommand{\bjdtdb}{\ensuremath{\rm {BJD_{TDB}}}}
\newcommand{\ecosw}{\ensuremath{e\cos{\omega_\star}}}
\newcommand{\esinw}{\ensuremath{e\sin{\omega_\star}}}
\newcommand{\msun}{\ensuremath{\,{\rm M}_\Sun}}
\newcommand{\rsun}{\ensuremath{\,{\rm R}_\Sun}}
\newcommand{\lsun}{\ensuremath{\,{\rm L}_\Sun}}
\newcommand{\mj}{\ensuremath{\,{\rm M}_{\rm J}}}
\newcommand{\rj}{\ensuremath{\,{\rm R}_{\rm J}}}
\newcommand{\fave}{\langle F \rangle}
\newcommand{\fluxcgs}{10$^9$ erg s$^{-1}$ cm$^{-2}$}
\newcommand{\kms}{\ensuremath{\,{\rm km~s^{-1}}}}

\usepackage{apjfonts}

\begin{document}
\title{KELT-6\lowercase{b}: A P\textasciitilde7.9 \lowercase{d} Hot Saturn Transiting a Metal-Poor Star with a Long-Period Companion\footnote{KELT\lowercase{ is a joint project of }T\lowercase{he }O\lowercase{hio }S\lowercase{tate }U\lowercase{niversity, }V\lowercase{anderbilt }U\lowercase{niversity, and }L\lowercase{ehigh }U\lowercase{niversity.}}}

\author{
Karen A.\ Collins\altaffilmark{1}, 
Jason D.\ Eastman\altaffilmark{2,3},
Thomas G.\ Beatty\altaffilmark{4},
Robert J.\ Siverd\altaffilmark{5},
B.\ Scott Gaudi\altaffilmark{4},
Joshua Pepper\altaffilmark{5,6},
John F.\ Kielkopf\altaffilmark{1},
John Asher Johnson\altaffilmark{7,8},
Andrew W.\ Howard\altaffilmark{9},
Debra A.\ Fischer\altaffilmark{10},
Mark Manner\altaffilmark{11,12},
Allyson Bieryla\altaffilmark{13},
David W.\ Latham\altaffilmark{13},
Benjamin J.\ Fulton\altaffilmark{9},
Joao Gregorio\altaffilmark{14},
Lars A.\ Buchhave\altaffilmark{15,16},
Eric L.\ N.\ Jensen\altaffilmark{17},
Keivan G.\ Stassun\altaffilmark{5,18},
Kaloyan Penev\altaffilmark{19},
Justin R.\ Crepp\altaffilmark{20},
Sasha Hinkley\altaffilmark{7,21},
Rachel A.\ Street\altaffilmark{2},
Phillip Cargile\altaffilmark{5},
Claude E.\ Mack\altaffilmark{5},
Thomas E.\ Oberst\altaffilmark{22},
Ryan L.\ Avril\altaffilmark{22},
Samuel N.\ Mellon\altaffilmark{22},
Kim K.\ McLeod\altaffilmark{23},
Matthew T.\ Penny\altaffilmark{4},
Robert P.\ Stefanik\altaffilmark{13},
Perry Berlind\altaffilmark{13},
Michael L.\ Calkins\altaffilmark{13},
Qingqing Mao\altaffilmark{5},
Alexander J.\ W.\ Richert\altaffilmark{24},
Darren L.\ DePoy\altaffilmark{25},
Gilbert A.\ Esquerdo\altaffilmark{13},
Andrew Gould\altaffilmark{4},
Jennifer L.\ Marshall\altaffilmark{25}, 
Ryan J.\ Oelkers\altaffilmark{25},
Richard W.\ Pogge\altaffilmark{4},
Mark Trueblood\altaffilmark{26},
and
Patricia Trueblood\altaffilmark{26}
}
\altaffiltext{1}{Department of Physics \& Astronomy, University of Louisville, Louisville, KY 40292, USA}
\altaffiltext{2}{Las Cumbres Observatory Global Telescope Network, 6740 Cortona Drive, Suite 102, Santa Barbara, CA 93117, USA}
\altaffiltext{3}{Department of Physics Broida Hall, University of California, Santa Barbara, CA 93106, USA}
\altaffiltext{4}{Department of Astronomy, The Ohio State University, 140 W.\ 18th Ave., Columbus, OH 43210, USA}
\altaffiltext{5}{Department of Physics and Astronomy, Vanderbilt University, Nashville, TN 37235, USA}
\altaffiltext{6}{Department of Physics, Lehigh University, Bethlehem, PA, 18015, USA}
\altaffiltext{7}{Department of Astrophysics, California Institute of Technology, Pasadena, CA 91125, USA}
\altaffiltext{8}{NASA Exoplanet Science Institute (NExScI), CIT Mail Code 100-22, 770 South Wilson Avenue, Pasadena, CA 91125, USA}
\altaffiltext{9}{Institute for Astronomy, University of Hawaii, 2680 Woodlawn Drive, Honolulu, HI 96822, USA}
\altaffiltext{10}{Department of Astronomy, Yale University, New Haven, Connecticut 06511, USA}
\altaffiltext{11}{Spot Observatory, Nunnelly, TN, USA}
\altaffiltext{12}{Montgomery Bell Academy, Nashville, TN, USA}
\altaffiltext{13}{Harvard-Smithsonian Center for Astrophysics, 60 Garden Street, Cambridge, MA 02138, USA}
\altaffiltext{14}{Atalaia Group \& Crow-Observatory, Portalegre, Portugal}
\altaffiltext{15}{Niels Bohr Institute, University of Copenhagen, Juliane Maries vej 30, 21S00 Copenhagen, Denmark}
\altaffiltext{16}{Centre for Star and Planet Formation, Geological Museum, {\O}ster Voldgade 5, 1350 Copenhagen, Denmark}
\altaffiltext{17}{Department of Physics and Astronomy, Swarthmore College, Swarthmore, PA 19081, USA}
\altaffiltext{18}{Department of Physics, Fisk University, Nashville, TN 37208, USA}
\altaffiltext{19}{Princeton University, Princeton, NJ, USA}
\altaffiltext{20}{Department of Physics, University of Notre Dame, 225 Nieuwland Science Hall, Notre Dame, IN 46556, USA}
\altaffiltext{21}{NSF Fellow}
\altaffiltext{22}{Westminster College, New Wilmington, PA, USA}
\altaffiltext{23}{Wellesley College, Wellesley, MA, USA}
\altaffiltext{24}{Department of Astronomy and Astrophysics, Pennsylvania State University, University Park, PA, USA}
\altaffiltext{25}{George P. and Cynthia Woods Mitchell Institute for Fundamental Physics and Astronomy, and Department of Physics \& Astronomy, Texas A~\&~M University, College Station, TX 77843-4242, USA}
\altaffiltext{26}{Winer Observatory, Sonoita, AZ 85637, USA}

\begin{abstract}
We report the discovery of KELT-6b, a mildly-inflated Saturn-mass
planet transiting a metal-poor host. The initial transit signal was
identified in KELT-North survey data, and the planetary nature of the
occulter was established using a combination of follow-up photometry,
high-resolution imaging, high-resolution spectroscopy, and precise
radial velocity measurements.  The fiducial model from a global
analysis including constraints from isochrones indicates that the
$V=10.38$ host star (BD+31 2447) is a mildly evolved, late-F star
with $\teff=6102\pm 43~{\rm K}$, $\loggstar=4.07_{-0.07}^{+0.04}$
and $\feh=-0.28\pm0.04$, with an inferred mass
$M_\star=1.09\pm0.04~\msun$ and radius
$R_\star=1.58_{-0.09}^{+0.16}~\rsun$. The planetary companion has mass
$M_{\rm P}=0.43\pm 0.05~\mjup$, radius
$R_{\rm P}=1.19_{-0.08}^{+0.13}~\rjup$, surface gravity
$\loggp=2.86_{-0.08}^{+0.06}$, and density
$\rho_{\rm P}=0.31_{-0.08}^{+0.07}~{\rm g~cm^{-3}}$.  The planet is on
an orbit with semimajor axis $a=0.079\pm0.001$AU and eccentricity
$e=0.22_{-0.10}^{+0.12}$, which is roughly consistent with circular, and has
ephemeris of $T_{\rm c}({\bjdtdb})=2456347.79679 \pm 0.00036$ and $P=7.845631 \pm
0.000046~{\rm d}$.  Equally plausible fits that employ empirical
constraints on the host star parameters rather than isochrones yield a
larger planet mass and radius by $\sim 4-7\%$.  KELT-6b has surface
gravity and incident flux similar to HD~209458b, but orbits a host
that is more metal poor than HD~209458 by $\sim 0.3$ dex.  Thus, the
KELT-6 system offers an opportunity to perform a comparative measurement of two similar
planets in similar environments around stars of very different
metallicities.  The precise radial velocity data also reveal an
acceleration indicative of a longer-period third body in the system,
although the companion is not detected in Keck adaptive optics
images.
\end{abstract}
\section{Introduction}

Ground-based surveys have discovered dozens of transiting exoplanets
around bright $(V<11)$ stars. Those discoveries are of considerable
importance because they enable cost-effective detailed measurements of
physical properties of extrasolar planets and their host stars (see
reviews by \citealt{winn2009,winn2010}). Discoveries of transiting exoplanets
that have characteristics similar to an already well-measured exoplanet, 
but that differ significantly in one aspect, are of particularly high 
importance because they enable comparative studies.

The high scientific value of transiting planet systems motivated the
first dedicated wide-field transit surveys, which have now produced a
large number of discoveries (TrES, \citealt{alon04}; XO,
\citealt{mc06}; HATNet, \citealt{bakos07}; SuperWASP, \citealt{cc07a},
QES, \citealt{al11}).  SuperWASP and HATNet have been especially
productive, with each survey discovering dozens of new transiting
planets. The space-based missions CoRoT \citep{bag03} and Kepler
\citep{boru10} have dramatically expanded the parameter space of
transit surveys, enabling the detection of transiting planets with sizes down to
that of the Earth and below, planets with periods of several years,
and planets orbiting host stars with a wider range of physical characteristics.

The Kilodegree Extremely Little Telescope-North (KELT-North) transit
survey \citep{pepper07} is designed to detect transiting planets
around bright stars. \citet{pepper03} designed the aperture, optical system, and exposure time
for KELT-North to provide better than 1\% RMS photometry
for stars with $8 < V < 10$.  That magnitude range represents the
brightness gap between comprehensive RV surveys and most other transit
surveys. The KELT-North telescope system was constructed using
commercial off-the-shelf equipment and has been collecting data since
September 2006.

The KELT-North survey has already announced three low-mass transiting
companions. KELT-1b \citep{siv12} is a highly inflated $27~M_{\rm J}$ brown
dwarf transiting a $V=10.7$ mid-F star. KELT-2Ab \citep{beatty12} is
a hot Jupiter transiting the bright $(V=8.77)$ primary star of a
binary system. KELT-3b \citep{pepper13} is a hot Jupiter planet
transiting a $V=9.8$ late-F star. The designations KELT-4 and KELT-5 
are currently reserved for two candidates in the confirmation phase.

Because KELT-North has focused on the same fields for an extended
length of time ($>6$ years), longer period ($P\ge 5~{\rm d}$)
planets are now detectable in the data. The large number of
observations of each field also enables the detection of smaller
planet-to-star radius ratios. In this paper we describe the discovery
and characterization of KELT-6b, a transiting mildly-inflated
Saturn-mass planet orbiting a $V=10.38$ metal-poor host BD+31 2447
(hereafter KELT-6). KELT-6b is currently the sixth longest period
exoplanet discovered by a ground-based transit
survey, after HAT-P-15b, HAT-P-17b, WASP-8b, WASP-59b, and 
WASP-84b\footnote{The Exoplanet Orbit Database (\citealt{wright11}; 
http://exoplanets.org/) lists four planets with longer periods as of November 5th, 2013. 
WASP-84b \citep{anderson2013} is not in the database
at the time of writing, but we include it here for completeness.}.
In several important aspects, KELT-6b resembles a metal-poor
analog of one of the most well-studied transiting
planets, HD~209458b \citep{charbonneau00,henry00}. Both hosts
have similar effective temperatures of $\sim 6100~{\rm K}$, although
KELT-6 is significantly more evolved and therefore has a 
larger radius.  On the other hand, KELT-6b has a substantially larger
orbit than HD 209458b.  As a result, the incident fluxes at both
planets are very similar.  In addition, the surface gravity of KELT-6b
differs from that of HD~209458b by only $\sim 20\%$. 

The discovery of KELT-6b offers an opportunity to perform a
comparative measurement of two similar planets in similar environments
around stars of very different metallicities. The comparison may, for example, 
elucidate the effect of bulk composition of the planet atmosphere on the cause of atmospheric
temperature inversions (e.g., \citealt{madhu2010}). 
In addition, host-star metallicity has been shown to affect the physical and orbital
properties of planets. In particular, there is a rough
correlation between metallicity and estimated core mass 
(\citealt{burrows2007}; \citealt{torres2008}; \citealt{sato2005}), and
there are indications of trends in the properties of planets with metallicity, which 
may signal the existence of multiple mechanisms for the formation and/or delivery 
of close-in giant planets (e.g., \citealt{ribas2007,dawson2013}).

\section{Discovery and Follow-up Observations}
\label{sec:obs}

We provide a brief summary of the KELT survey data reduction process in \S \ref{sec:keltobs};
for more details, see \S2 of \citet{siv12}.

\subsection{KELT-North Observations and Photometry\label{sec:keltobs}}

KELT-6 is in KELT-North survey field 08, which is centered on
($\alpha=13^h38^m28^s.25$, $\delta=+31\arcdeg41\arcmin12\farcs67$; J2000).  We
monitored field 08 from December 2006, to June 2011, collecting a
total of 7359 observations. We reduced the raw survey data using a
custom implementation of the ISIS image subtraction package
\citep{al98, al00}, combined with point-spread fitting photometry
using DAOPHOT \citep{stet87}. Using proper motions from the Tycho-2
catalog \citep{hog00} and J and H magnitudes from 2MASS (\citealt{skrutskie06}; \citealt{cutri03}),
we implemented a reduced proper motion cut \citep{gm03}
based on the specific implementation of \citet{cc07b}, in order to select likely dwarf and subgiant stars
within the field for further post-processing and analysis. We applied
the trend filtering algorithm \citep[TFA;][]{k05} to each remaining
light curve to remove systematic noise, followed by a search for
transit signals using the box-fitting least squares algorithm
\citep[BLS;][]{k02}. For both TFA and BLS we used the versions found
in the VARTOOLS package \citep{hart08}.

\begin{figure}[ht]
\begin{center}
\includegraphics[scale=0.43]{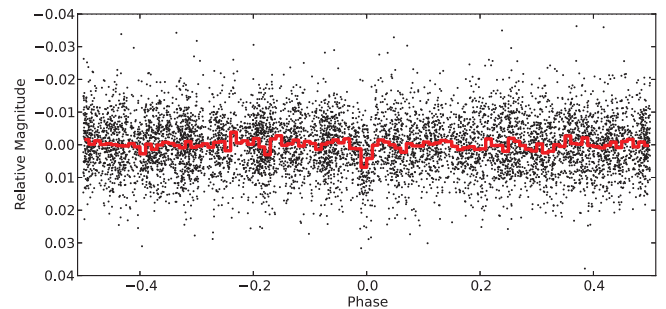}
\caption{Discovery light curve of KELT-6 from the KELT-North telescope.  The light curve contains 7359 observations spanning 4.5 years, phase-folded to the orbital period of 7.8457 days.  The solid red line represents the same data binned at $\sim$2-hour intervals after phase-folding. \label{fig:kelt_lc}}
\end{center}
\end{figure}

One of the candidates from field 08 was star BD+31 2447 / TYC~2532-556-1,
 located at ($\alpha=13^h03^m55^s.65$,
$\delta=+30\arcdeg38\arcmin24\farcs3$; J2000).  The star has
Tycho magnitudes $B_{\rm T} = 10.736 \pm 0.048$ and $V_{\rm T} = 10.294 \pm 0.050$
\citep{hog00}, and passed our initial selection cuts.  The discovery
light curve of KELT-6 is shown in Figure \ref{fig:kelt_lc}.  We
observed a transit-like feature at a period of 7.8457 days, with a
depth of about 5 mmag.

\subsection{Radial-Velocity Observations\label{sec:RV}}

After KELT-6 was selected as a candidate, we conducted radial-velocity
(RV) observations to identify possible false-positive signatures and
to determine the RV orbit.  We obtained data using the Tillinghast
Reflector Echelle Spectrograph\footnote{http://tdc-www.harvard.edu/instruments/tres/}
\citep[TRES;][]{tres}, on the 1.5m Tillinghast Reflector at the Fred
L. Whipple Observatory (FLWO) at Mt. Hopkins, AZ.  We observed KELT-6
three times with TRES over three months, from UT 2012-04-12 to UT
2012-07-09.  The spectra have a resolving power of R=44,000, and were
extracted following the procedures described by \citet{bu10}. These 
three initial TRES single-order absolute RVs are listed in Table \ref{tab:rv1} and are consistent with no RV
variations to within the errors, ruling out some classes of astrophysical
false positives. However, the TRES RV uncertainties
are large enough to still allow for a low-mass companion at the
$\sim 7.8$d period of the KELT-North candidate signal, and on
that basis we chose to continue with photometric follow-up.  Note
that due to the relatively large uncertainties, we chose not
to include these TRES velocities in the final global analysis
described in \S \ref{sec:exofast}.

On UT 2012-06-26, we obtained high precision KELT-6 follow-up
photometry of the final third of a predicted transit and
detected an apparent shallow egress (see \S \ref{sec:phot}). Based on
that detection and the lack of RV variations in the TRES data, we
decided to pursue higher-precision RV data.

\begin{figure}
\begin{center}
\includegraphics[scale=0.9,trim=0mm 0mm 0mm 0mm,clip]{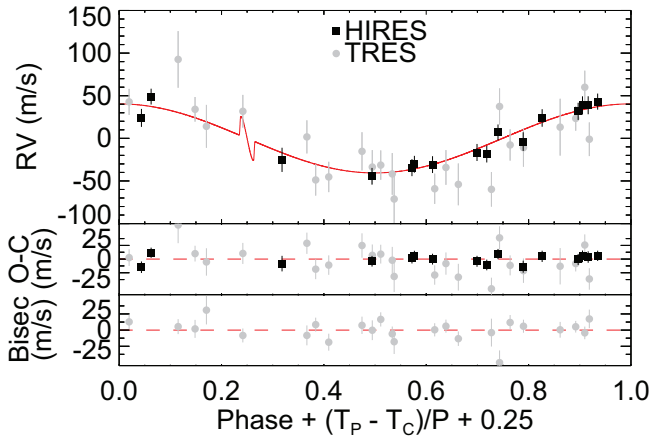}
\caption{HIRES and TRES relative radial velocity measurements of KELT-6. {\it Top panel:} Relative RV observations phased to our fiducial orbital model (see \S \ref{sec:global_fits}) which is fit to the HIRES data only with eccentricity and the RV linear trend as free parameters. The fiducial model is shown as a solid red line. The predicted Rossiter-McLaughlin effect incorporates an assumption that $\lambda=0$ (i.e. that the projected spin-obit alignment of the system is 0 degrees).  HIRES observations are shown as black squares and the error bars are scaled according to the method described in \S \ref{sec:exofast}. TRES observations are shown as gray circles with unrescaled errors. These data were not used in the fit, but are simply phased to the period of the fiducial model, and shifted by a constant offset that minimizes the $\chisq$ of the data from the fiducial model. {\it Middle panel:} Residuals of the RV observations to the fiducial fit. The RMS of the HIRES RV residuals is 8.0 m s$^{-1}$. {\it Bottom panel:} Bisector spans of the TRES spectra. \label{fig:rv}}
\end{center}
\end{figure}

Using the High Resolution Echelle Spectrometer (HIRES) instrument
\citep{vogt94} on the Keck I telescope located on Mauna Kea, Hawaii,
we obtained 16 exposures between UT 2012-08-24 and UT 2013-02-21 with
an iodine cell, plus a single iodine-free template spectrum. The
absolute and precise relative RV measurements are listed in Table \ref{tab:rv1}, and Figure
\ref{fig:rv} shows the HIRES relative RV data phased to the orbit fit with a
linear trend of $\dot{\gamma}=-0.239$ m s$^{-1}$ day$^{-1}$ (see \S
\ref{sec:exofast}) removed, along with the residuals to the model
fit.

The HIRES radial velocity observations were made using the standard setup of
the California Planet Search (CPS) program
\citep{johnson10,howard11}. A pyrex cell containing gaseous iodine is
placed in front of the spectrometer entrance slit, which imprints a
dense set of molecular iodine lines on each stellar spectrum. The
iodine lines provide a calibration of the instrumental profile as well
as a precise measure of the wavelength scale at the time of observation
\citep{mb92}. We measured the relative stellar radial velocities using
the forward-modeling scheme of \citet{butler96} with improvements made
over the years. We measured the absolute RVs using the methods
of \citet{chubak2012}.

The PSF varies quite dramatically in the slit-fed HIRES instrument
simply from guiding and spectrometer focus variations.  Since line
asymmetries due to instrumental and stellar sources cannot be easily
distinguished, we do not attempt to measure bisector spans for the
HIRES observations.

We also obtained five RV measurements between UT 2013-02-01 and UT
2013-02-15 using the Hobby-Eberly Telescope (HET). However, these data
were taken without an iodine cell for wavelength reference, and as a
result the uncertainties are >6 km s$^{-1}$, so we do not list them in
the RV table or use them in the global fit analysis in \S
\ref{sec:exofast}.

Finally, 21 additional TRES RVs were obtained and reduced using
multi-order analysis after most of the global analysis had been
completed. The full TRES RV dataset is listed in Table \ref{tab:rv1}
and contains measurements from 24 different nights between UT 2012-04-12 
and UT 2013-07-31, with typical relative uncertainties of 20 m~s$^{-1}$. 
Although we do not use the TRES RV data in our global fit analysis (see \S \ref{sec:global_fits}), 
we note that these data independently
confirm both the amplitude of the KELT-6b RV variations (see Figure \ref{fig:rv}) and the linear
trend of the fiducial global fit (see \S \ref{sec:tertiary}), albeit with larger uncertainties due to the
somewhat worse precision than the Keck data. Bisector spans were
calculated from the TRES spectra following \citet{torres2007} and are used 
in \S \ref{sec:falsepos} as part of the false positive analysis. The bisector spans 
are listed in Table \ref{tab:rv1}, and shown in the bottom panel of Figure \ref{fig:rv} 
phased to the orbital fit.

\begin{center}
\begin{table}
\caption{Radial Velocity and Bisector Span Variation Measurements of KELT-6\label{tab:rv1}}
{\setlength{\tabcolsep}{0.65em}
\begin{tabular}{lrrrrrc}
\tableline
\vspace{-0.1in}\\
\multicolumn{1}{c}{\bjdtdb} & \multicolumn{1}{c}{Abs} & \multicolumn{1}{c}{Rel} & \multicolumn{1}{c}{Rel} & \multicolumn{1}{c}{BS\tablenotemark{d}} & \multicolumn{1}{c}{$\sigma_{BS}$\tablenotemark{e}} & \multicolumn{1}{c}{Source} \\
\multicolumn{1}{c}{} & \multicolumn{1}{c}{RV\tablenotemark{a}} & \multicolumn{1}{c}{RV\tablenotemark{b}} & \multicolumn{1}{c}{$\sigma_{RV}$\tablenotemark{c}} & \multicolumn{1}{c}{} & \multicolumn{1}{c}{} &  \\
%\multicolumn{1}{c}{} & \multicolumn{1}{c}{(m s$^{-1}$)} & \multicolumn{1}{c}{(m s$^{-1}$)} & \multicolumn{1}{c}{(m s$^{-1}$)} & %\multicolumn{1}{c}{(m s$^{-1}$)} &  \\
\tableline
\vspace{-0.1in}\\
2456029.869867  &  1085   & 27.02 & 20.27 & -3.9 & 21.6 &  TRES \\
2456114.681684  &  1188  & -5.96 & 28.33 & -18.0 & 19.0 & TRES \\
2456117.673406  &  1166   & 63.43 & 19.73 & 17.3 & 14.4 & TRES \\
\tableline
\vspace{-0.1in}\\
2456163.733467  & 1300  & 19.38  &  3.90   &  - & - &  HIRES \\
2456164.729771  & 1216  &  61.75  &  3.76   &  - & - &  HIRES \\
2456165.727060  & 1202  &  46.70  &  3.76   &  - & - &  HIRES \\
2456172.720200  & 1209  & 64.03  &  3.36   &  - & - &  HIRES \\
2456173.720971  & 1163 &  69.62  &  3.31   &  - & - &  HIRES \\ 
2456175.727980  & 1081 & -5.05  &  5.06   &  - & - &  HIRES \\
2456177.719323  & 1161 &  -14.71 &   3.56   &  - & - &  HIRES \\
2456178.716548  & 1000 &   2.82  &  3.57   &  - & - &  HIRES \\
2456179.715151  & 1214 &  43.05  &  3.67   &  - & - &  HIRES \\
2456290.173359  & 1311 &  29.64  &  3.81   &  - & - &  HIRES \\
2456318.978815  & 1132 &  -46.73  &  3.50   &  - & - &  HIRES \\
2456320.088103  & 1319 & -35.17  &  3.48   &  - & - &  HIRES \\
2456326.175000  & 1185 &  -62.66  &  3.50   &  - & - &  HIRES \\
2456327.103147  & 1340 &  -49.24  &  3.52   &  - & - &  HIRES \\
2456328.106726  & 1333 &  -11.70  &  3.32   &  - & - &  HIRES \\
2456345.026477  & 1291 &  8.71  &  3.58   &  - & - &  HIRES \\
\tableline
\vspace{-0.1in}\\
2456443.707006	& 1109	 & -33.79   &    22.37 & 7.4 & 13.4 & TRES \\
2456450.707562	& 1125	 & -18.64    &   19.40 & -8.2 & 15.4 & TRES \\
2456451.710411	& 1021	 & -54.28    &   19.92 &  -0.2 & 15.3 & TRES \\
2456452.667049	& 1022  & -79.97   &    17.75 &  0.5 & 10.2 & TRES \\
2456453.661056	& 1044  &  16.31   &    21.20 &  -50.3 & 18.9 & TRES \\
2456458.688259	& 1015  &  -71.18   &    18.35 &  8.5 & 10.8 & TRES \\
2456459.685632	& 1147  &  -54.01   &    17.19 &  16.6 & 11.0 & TRES \\
2456460.684562	& 1022	 &   -57.04   &    20.36 &  6.1 & 10.1 & TRES \\
2456461.669084	& 1127	 &  -30.88    &   18.32 &  11.8 & 11.6 & TRES \\
2456462.673881	& 1181	  &    0.00   &    14.42 &  5.2 & 9.0 & TRES \\
2456463.673604	& 1150	  &   19.16   &    15.25 &  12.8 & 11.2 & TRES \\
2456464.684882	& 1141	  &  10.16   &    14.42 &  1.9 & 13.9 & TRES \\
2456466.733703	& 1045	  &  -69.70   &    17.79 &  -18.7 & 13.2 & TRES \\
2456467.707369	& 1089	  &  -66.40   &    24.43 &  -5.8 & 11.4 & TRES \\
2456468.720609	& 1071	  &  -78.93   &    24.46 &  -13.3 & 11.2 & TRES \\
2456469.719214	& 1080   &  -36.11   &    22.32 &  6.0 & 10.1 & TRES \\
2456470.658849	& 1168	  &   34.53   &    19.41 &  -4.3 & 9.8 & TRES \\
2456472.704518	& 1211   &  -11.86   &    25.12 &  30.9 & 22.1 & TRES \\
2456501.663546	& 1628   &  -20.17   &    33.33  & 0.5 & 12.2 &  TRES \\
2456503.653861	& 1697	  &   58.79   &    33.22 &  5.4 & 11.6 & TRES \\
2456504.645862	& 1577  &  -2.35   &    19.21 &  -8.3 & 10.8 & TRES \\
\tableline
\end{tabular}}
\tablecomments{Absolute RVs are on the IAU scale. Based on extensive observations of radial velocity reference stars, the native absolute velocity scale of TRES has been transformed to the IAU absolute velocity scale by subtracting 610 m s$^{-1}$. The absolute RV error is 100 m~s$^{-1}$ and is dominated by the long-term RMS for velocity standard stars. The relative RV values reported are on the native system for each instrument and cannot be directly compared to values from a different instrument. The bisector spans (BS) from the TRES spectra are computed as described in the text.}
\tablenotetext{a}{absolute RVs (m s$^{-1}$)}
\tablenotetext{b}{relative RVs (m s$^{-1}$)}
\tablenotetext{c}{unrescaled relative RV errors (m s$^{-1}$)}
\tablenotetext{d}{spectral line bisector spans (m s$^{-1}$)}
\tablenotetext{e}{spectral line bisector span errors (m s$^{-1}$)}
\end{table}
\end{center}

\vspace{1in}
\subsection{Follow-up Time-Series Photometry} \label{sec:phot}

\begin{figure}[ht]
\begin{center}
\includegraphics[scale=0.75,trim=0mm 0mm 0mm 0mm,clip]{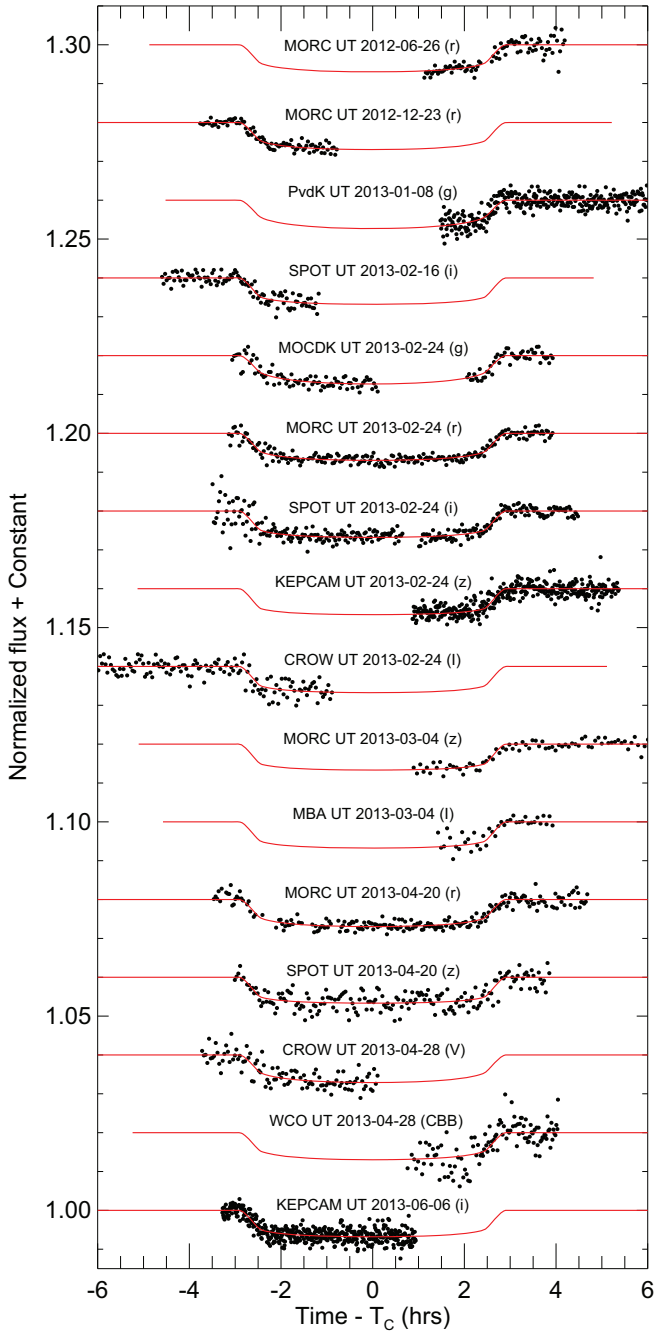}
\caption{Follow-up transit photometry of KELT-6. The red overplotted lines are the best fit transit model from global fit 6 described in \S \ref{sec:global_fits} and summarized in Table \ref{tab:multipars}. The transit times are shown in Table \ref{tab:ttvs}. The labels are as follows:
MORC=University of Louisville Moore Observatory 0.6 m RCOS Telescope;
PvdKO=Peter van de Kamp Observatory 0.6 m RCOS Telescope;
SPOT=Spot Observatory 0.6 m RCOS Telescope;
MOCDK=University of Louisville Moore Observatory PlaneWave 0.5 m CDK Telescope;
KEPCAM=Keplercam at the Fred Lawrence Whipple Observatory 1.2 m Telescope;
CROW=Canela's Robotic Observatory 0.3 m LX200 Telescope;
MBA=Montgomery Bell Academy 0.6 m PlaneWave CDK Telescope;
WCO=Westminster College Observatory 0.35 m C14 Telescope.
\label{fig:followup_lcs}}
\end{center}
\end{figure}

\begin{figure}
\begin{center}
\includegraphics[scale=0.9]{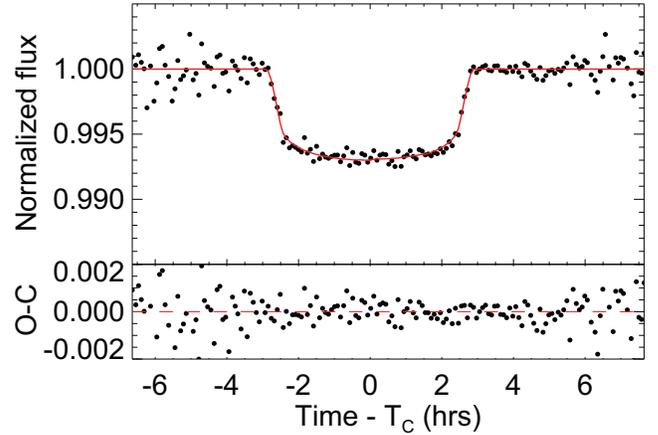}
\caption{{\it Top panel:} All follow-up light curves from Figure \ref{fig:followup_lcs} (except the WCO light curve - see text), combined and binned in 5 minute intervals.  This light curve is not used for analysis, but rather to show the best combined behavior of the transit.  The red curve shows the 15 transit models from global fit 6 described in Table \ref{tab:multipars} for each of the individual fits combined and binned in 5 minute intervals the same way as the data, with the model points connected. {\it Bottom panel:} The residuals of the binned light curve from the binned model in the top panel.\label{fig:bestlc}}
\end{center}
\end{figure}

We acquired follow-up time-series photometry of KELT-6 to check for other types of false positives and to better determine the transit shape.  To schedule follow-up photometry, we used the {\tt Tapir} software package \citep{jensen2013}.  We obtained 16 partial or full primary transits in multiple bands between June 2012 and June 2013. The transit duration ($>5.5$ hours) and orbital period ($>7.8$ days) are long, so opportunities to observe full transits are rare. Figure \ref{fig:followup_lcs} shows all the primary transit follow-up light curves assembled. A summary of the follow-up photometric observations is shown in Table \ref{tab:photobs}. We find consistent $R_{\rm P}/R_\star$ ratios in all light curves, which include observations in the $g$, $r$, $i$, $z$, $V$, $I$, and $CBB$ filters\footnote{In all references to SDSS filters in this paper, we use the unprimed notation to denote generic SDSS-like filters, which in practice are often labeled with the primed notation. CBB denotes the Astrodon clear with blue block filter which starts transmitting near 500 nm and continues to transmit into the near-infrared.}, helping to rule out false positives due to blended eclipsing binaries.  Figure \ref{fig:bestlc} shows all primary transit follow-up light curves from Figure \ref{fig:followup_lcs} (except the WCO light curve which contains significant residual systematics after detrending), combined and binned in 5 minute intervals. This combined and binned light curve is not used for analysis, but rather to show the best combined behavior of the transit. We also observed KELT-6 near the uncertain time of secondary transit on five different epochs (see \S \ref{sec:falsepos}). 

Unless otherwise noted, all photometric follow-up observations were reduced with the AstroImageJ (AIJ) package\footnote{http://www.astro.louisville.edu/software/astroimagej/} (K.~A.~Collins \& J.~F.~Kielkopf 2014, in preparation). AIJ is a general purpose image processing package, but is optimized for processing time-series astronomical image sequences. It is open source software written in Java and is compatible with all computing platforms commonly used to process astronomical data. AIJ is a graphical user interface driven package that provides an interactive multi-image display interface, CCD image calibration (bias, dark, flat-field, and non-linearity correction), astronomical time and coordinate calculations, multi-aperture differential photometry, multi-dataset plotting, and interactive light curve detrending. It can be operated in combination with any camera control software to reduce data and plot differential light curves in real time, or can be used in standard mode to post process data.

Also unless otherwise noted, calibration of all photometric follow-up observations included bias and dark subtraction followed by flat-field correction. Calibration of the MORC data also included a correction for CCD non-linearity. Differential photometry was performed on the calibrated images using a circular aperture.

We observed three complete and three partial transits of KELT-6 using two telescopes at Moore Observatory, operated by the University of Louisville.  The 0.6 m RCOS telescope with an Apogee U16M 4K $\times$ 4K CCD, giving a 26' $\times$ 26' field of view and 0.39 arcseconds pixel$^{-1}$, was used to observe the $r$ egress on UT 2012-06-26, the $r$ ingress on UT 2012-12-23, the full $r$ transit on UT 2013-02-24, the $z$ egress on UT 2013-03-04, and the full $r$ transit on UT 2013-04-20. The 0.6 m was also used to observe near the time of secondary transit on UT 2013-04-16 in $z$. The Planewave Instruments 0.5 m CDK telescope with an Apogee U16M 4K $\times$ 4K CCD, giving a 37' $\times$ 37' field of view and 0.54 arcseconds pixel$^{-1}$, was used to observe most of a transit in $g$ on UT 2013-02-24. The gap in the data is due to a meridian flip. 

We observed an egress in $g$ at Swarthmore College's Peter van de Kamp Observatory on UT 2013-01-08.  The observatory uses a 0.6 m RCOS telescope with an Apogee U16M 4K $\times$ 4K CCD, giving a 26' $\times$ 26' field of view. Using 2 $\times$ 2 binning, it has 0.76 arcseconds pixel$^{-1}$.  

We observed one partial and two full transits at Spot Observatory.  The observatory uses a 0.6 m RCOS telescope with an SBIG STX 16803 4K $\times$ 4K CCD, giving a 26' $\times$ 26' field of view and 0.39 arcseconds pixel$^{-1}$. An ingress in $i$ was observed on UT 2013-02-16, and full transits were observed on UT 2013-02-24 in $i$ and UT 2013-04-20 in $z$. We also observed near the time of secondary transit on UT 2013-02-20 in $z$

We observed an egress in $z$ on UT 2013-02-24 and an ingress in $i$ on UT 2013-06-06 with KeplerCam on the 1.2 m telescope at FLWO. KeplerCam has a single 4K $\times$ 4K Fairchild CCD with 0.366 arcseconds pixel$^{-1}$, and a field of view of 23.1' $\times$ 23.1'. We also observed near the time of secondary transit on UT 2013-02-28, UT 2013-04-08, and UT 2013-04-24 in $z$.

We observed one full and one partial transit at Montgomery Bell Academy (MBA) Long Mountain Observatory.  The observatory uses a PlaneWave Instruments 0.6 m CDK telescope with an SBIG STL 11002 4008 $\times$ 2672 CCD, giving a 30' $\times$ 20' field of view and 0.45 arcseconds pixel$^{-1}$. A full transit was observed in $V$ on UT 2013-02-24. However, the resulting light curve had large systematics that we were unable to adequately remove. Since the same transit epoch was observed by both Moore Observatory telescopes in overlapping filter bands, these data added no new information to the analysis and was not included in the global fit described in \S \ref{sec:exofast}. An egress in $I$ was observed on UT 2013-03-04, and observations near the time of secondary transit were collected in $z$ on UT 2013-02-20.

We observed two partial transits at Canela's Robotic Observatory (CROW) in Portugal. The observations were obtained using a 0.3 m LX200 telescope with an SBIG ST-8XME 1530 $\times$ 1020 CCD, giving a 28' $\times$ 19' field of view and 1.11 arcseconds pixel$^{-1}$. An ingress was observed in $I_c$ on UT 2013-02-24, and an ingress was observed in $V$ on UT 2013-04-28.

We observed a partial transit at Westminster College Observatory (WCO) in Pennsylvania. The observations were obtained using a Celestron 0.35 m C14 telescope with an SBIG STL-6303E 3072 $\times$ 2048 CCD, giving a 24' $\times$ 16' field of view and 1.4 arcseconds pixel$^{-1}$ at 3 x 3 pixel binning. An egress was observed using an Astrodon Clear with Blue Blocking ($CBB$) filter on UT 2013-04-28. 

We observed near the time of secondary transit on UT 2013-04-08 and UT 2013-04-24 using the 1.0 m telescope at the ELP node of the Las Cumbres Observatory Global Telescope (LCOGT) network at McDonald observatory in Texas \citep{brown13}. The observations were obtained in the Pan-STARRS-Z band with an SBIG STX-16803 4096 $\times$ 4096 CCD, giving a 15.8' $\times$ 15.8' field of view and 0.464 arcseconds pixel$^{-1}$ (2$\times$2 binning). The ELP data were processed using the pipeline discussed in \citet{brown13}.

\begin{table}
\caption{Summary of Photometric Observations\label{tab:photobs}}
{\setlength{\tabcolsep}{0.30em}
\begin{tabular}{lcrcrccc}
\tableline
\multicolumn{1}{l}{Telescope} & \multicolumn{1}{c}{UT} & \multicolumn{1}{c}{\#} & \multicolumn{1}{c}{Band} & \multicolumn{1}{c}{Cycle\tablenotemark{a}} & \multicolumn{1}{c}{RMS\tablenotemark{b}} & \multicolumn{1}{c}{PNR\tablenotemark{c}} & \multicolumn{1}{c}{Detrend} \\ 
\multicolumn{1}{l}{ } & \multicolumn{1}{c}{Date} & \multicolumn{1}{c}{Obs} & \multicolumn{1}{c}{ } & \multicolumn{1}{c}{(sec)} & \multicolumn{1}{c}{($10^{-3}$)}  & \multicolumn{1}{c}{($\frac{10^{-3}}{minute}$)} & \multicolumn{1}{c}{Variables}\\
\vspace{-0.1in}\\
\tableline
\sidehead{Primary:}
MORC & 2012-06-26 & 87 & $r$ & 119 & 1.6 & 2.3 & AM\\
MORC  & 2012-12-23 & 91 & $r$ & 119 & 0.8 & 1.1 & AM\\
PvdKO   & 2013-01-08  & 315 & $g$ & 52 & 1.7 & 1.6 & AM,PK\\
SPOT  & 2013-02-16  & 104 & $i$ & 119 & 1.4 & 2.0 & AM,TM\\
MOCDK  & 2013-02-24  & 131 & $g$ & 141 & 1.3 & 2.0 & AM,MF,SK\\
MORC  & 2013-02-24  & 212 & $r$ & 119 & 1.0 & 1.4 & AM,FW\\
SPOT  & 2013-02-24  & 278 & $i$ & 99 & 1.8 & 2.3 & AM,MF\\
KEPCAM  & 2013-02-24  & 361 & $z$ & 45 & 1.6 & 1.4 & AM,SK\\
CROW  & 2013-02-24  & 148 & $I$ & 142 & 1.9 & 2.9 & AM,PK,TM\\
MORC  & 2013-03-04  & 95 & $z$ & 259 & 1.0 & 2.1 & AM,TM\\
MBA  & 2013-03-04  & 39 & $I$ & 236 & 1.5 & 3.0 & AM\\
MORC  & 2013-04-20  & 212 & $r$ & 119 & 1.2 & 1.7 & AM\\
SPOT  & 2013-04-20  & 179 & $z$ & 139 & 2.0 & 3.0 & AM,TM\\
CROW  & 2013-04-28  & 102 & $V$ & 135 & 2.1 & 3.2 & AM,MF\\
WCO  & 2013-04-28  & 114 & $CBB$ & 105 & 3.0 & 4.0 & AM,TM\\
KEPCAM  & 2013-06-06  & 441 & $i$ & 35 & 1.6 & 1.2 & AM\\
\tableline
\sidehead{Secondary:}
MBA  & 2013-02-20  & 58 & $z$ & 236 & 1.5 & 3.0 & AM\\
SPOT  & 2013-02-20 & 47 & $z$ & 259 & 1.3 & 2.7 & AM\\  
KEPCAM  & 2013-02-28 & 757 & $z$ & 35 & 2.0 & 1.5 & AM\\ 
KEPCAM  & 2013-04-08 & 324 & $z$ & 45 & 1.7 & 1.5 & AM,XY\\ 
ELP  & 2013-04-08 & 162 & $PS$-$Z$ & 90 & 1.9 & 2.3 & AM,XY\\ 
MORC  & 2013-04-16 & 72 & $z$ & 259 & 1.4 & 2.9 & AM\\ 
ELP  & 2013-04-24 & 204 & $PS$-$Z$ & 90 & 2.5 & 3.1 & AM,XY\\ 
KEPCAM  & 2013-04-24 & 701 & $z$ & 35 & 2.4 & 1.8 & AM\\ 
\tableline
\end{tabular}}
\tablecomments{
MORC=University of Louisville Moore Observatory 0.6 m RCOS Telescope;
PvdKO=Peter van de Kamp Observatory 0.6 m RCOS Telescope; 
SPOT=Spot Observatory 0.6 m RCOS Telescope; 
MOCDK=University of Louisville Moore Observatory PlaneWave 0.5 m CDK Telescope; 
KEPCAM=Keplercam at the Fred Lawrence Whipple Observatory 1.2 m Telescope; 
CROW=Canela's Robotic Observatory 0.3 m LX200 Telescope; 
MBA=Montgomery Bell Academy 0.6 m PlaneWave CDK Telescope; 
WCO=Westminster College Observatory 0.35 m C14 Telescope;
ELP=McDonald 1.0 m Telescope (Las Cumbres Observatory Global Telescope Network); 
AM=airmass;
PK=peak count in aperture;
TM=time;
MF=meridian flip;
SK=sky background;
FW=average FWHM in image;
XY=detector x,y coordinates of target star centroid;
$PS$-$Z$=Pan-STARRS-Z.
}
\tablenotetext{a}{Cycle time in seconds, calculated as the mean of exposure time plus dead time during periods of back-to-back exposures.}
\tablenotetext{b}{RMS of residuals from the best fit model in units of $10^{-3}$ .}
\tablenotetext{c}{Photometric noise rate in units of $10^{-3}$ minute$^{-1}$, calculated as RMS/$\sqrt{\Gamma}$, where RMS is the scatter in the light curve residuals and $\Gamma$ is the mean number of cycles (exposure time and dead time) per minute during periods of back-to-back exposures (adapted from \citealt{fulton2011}).}
\end{table}

\subsection{Adaptive Optics Observations\label{sec:ao}}

\begin{figure}
\begin{center}
\fbox{\includegraphics[scale=0.75]{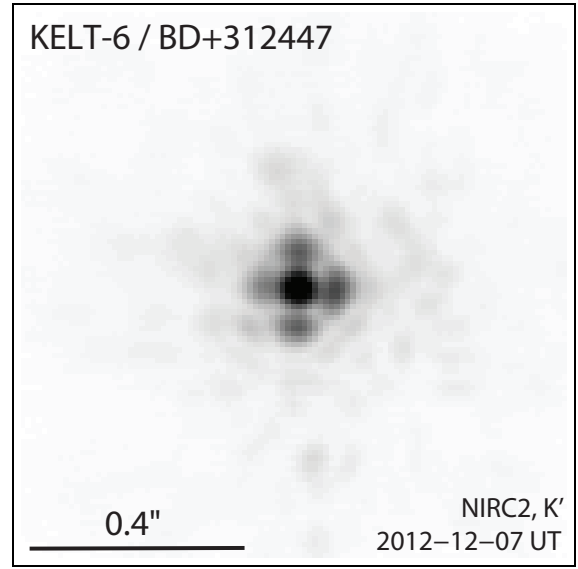}}
\caption{Keck adaptive optics image of KELT-6 taken with NIRC2 in the K' filter. The image is displayed on a negative square-root intensity scale to emphasize the surrounding regions. North is up, and east is left.\label{fig:ao}}
\end{center}
\end{figure}

We obtained adaptive optics (AO) imaging using NIRC2 
(instrument PI: Keith Matthews) at 
Keck on UT 2012-12-07. The AO imaging places limits on the existence 
of nearby eclipsing binaries that could be blended with the primary star 
KELT-6 at the resolution of the KELT and follow-up data, thereby causing 
a false positive planet detection.  In addition, it places limits on any 
nearby blended source that could contribute to the total flux, and
thereby result in an underestimate of the transit depth and thus 
planet radius in the global fit presented in \S \ref{sec:exofast}. 
Our observations consist of dithered frames taken 
with the $K'$ filter. We used the narrow camera setting to provide fine 
spatial sampling of the stellar point-spread function, and used KELT-6 
as its own on-axis natural guide star. The total on-source integration 
time was 225 seconds.  The resulting image is shown in Figure \ref{fig:ao}.

\begin{figure}
\begin{center}
\includegraphics[scale=0.34]{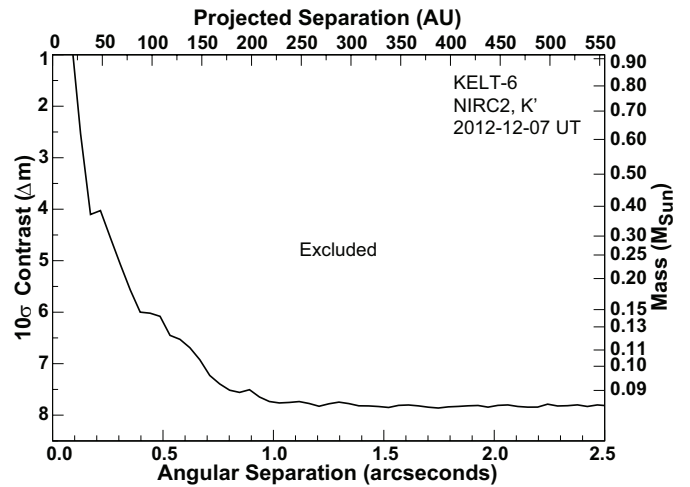}
\caption{Contrast sensitivity derived from the Keck adaptive optics image of KELT-6 shown in Figure \ref{fig:ao}. The $10\sigma$ contrast in $\Delta$magnitude is plotted against angular separation in arcseconds. The scale on top shows projected separation in AU for a distance of 222 pc (see Table \ref{tab:hostprops}). The scale on the right side of the plot estimates the mass in units of $\msun$ at a given contrast, estimated using the \citet{baraffe1998} models.  We can exclude companions beyond a distance of 0.5 arcseconds (111 AU) from KELT-6 down to a magnitude difference of 6.0 magnitudes at $10\sigma$. \label{fig:contrast}}
\end{center}
\end{figure}

We find no significant detection of off-axis sources in the immediate
vicinity of KELT-6.  We note that there are some conspicuous sources
at the threshold of detection.  However, without an image in a
different filter, we are unable to determine if the position of these
sources are wavelength dependent, which would indicate that they are
speckles rather than real sources.  Nevertheless, we can still place a
conservative upper limit on any real sources based on the contrast
sensitivity.  Figure \ref{fig:contrast} shows the $10\sigma$ contrast
sensitivity (in $\Delta$magnitude) versus angular separation computed
from Figure \ref{fig:ao} using a three-point dither pattern to build signal
and subtract sky-background (see \citealt{crepp2012}). The top scale in Figure \ref{fig:contrast} shows 
projected separation in AU for a distance of 222 pc (see Table \ref{tab:hostprops}). 
The scale on the right side of the plot estimates the mass in units of $\msun$ at 
a given contrast, estimated using the \citet{baraffe1998} models. We can 
exclude companions beyond a distance of 0.5 arcseconds (111 AU) from KELT-6 
down to a magnitude difference of 6.0 magnitudes at $10\sigma$.

\section{Host Star Properties}
\label{sec:star}

\subsection{Properties from the Literature\label{sec:star_props}}

Table \ref{tab:hostprops} lists various properties and measurements of KELT-6 collected from the literature and derived in this work. The data from the literature include FUV and NUV fluxes from GALEX \citep{galex05}, $B-V$ color from \citet{harris64}, optical fluxes in the $B_{\rm T}$ and $V_{\rm T}$ passbands from the Tycho-2 catalog \citep{hog00}, $V$ and $I_{\rm C}$ from The Amateur Sky Survey (TASS;  \citealt{tass2000}), near-infrared (IR) fluxes in the $J$, $H$ and $K_{\rm S}$ passbands from the 2MASS Point Source Catalog (\citealt{skrutskie06}; \citealt{cutri03}), near- and mid-IR fluxes in three WISE passbands (\citealt{wright10}; \citealt{cutri12}), and proper motions from the NOMAD catalog \citep{zacharias04}.

\subsection{Spectroscopic Analysis\label{sec:spec_params}}

We use both the TRES and HIRES spectra to derive the stellar
properties of KELT-6.  To analyze the TRES spectra, we use the
Spectral Parameter Classification (SPC) procedure version 2.2
\citep{bu12} with $\teff$, $\loggstar$, [m/H], and $\vsini$ as
free parameters. 
Since each of the 24 TRES spectra yielded similar results, we took the mean value for
each stellar parameter. The uncertainties are dominated by systematic rather
than statistical errors, so we adopt the mean error for each parameter. The results are: $\teff = 6098 \pm 50$
K, $\loggstar = 3.83 \pm 0.10$, [m/H] = $-0.34 \pm 0.08$, and $\vsini
= 6.7 \pm 0.5~{\rm km~s^{-1}}$, giving the star an inferred spectral type
of F8.  

To analyze the HIRES spectra, we use spectral synthesis modeling with
Spectroscopy Made Easy (SME, \citealt{vp96}, \citealt{vf05}).
The free parameters for the model included $\teff$, $\vsini$, $\loggstar$,
and $\feh$. The microturbulent velocity was fixed to 0.85 $\rm km
s^{-1}$ in this model and the macroturbulent velocity was specified as
a function of effective temperature \citep{vf05}.  After the first
model was generated, two other iterations were run with temperature
offsets of $\pm 100$ K from the model temperature to evaluate degeneracy 
between the model parameters.  If the RMS for these new fit parameters
relative to the original model values exceeds the uncertainties on the
original model values estimated using the error analysis of
\citet{vf05}, then these larger uncertainties are adopted.  However,
in this case, the fits starting with the temperature offsets settled
on values very close to those found using the original
model, differing by much less than the estimated uncertainties on the
original model values.  Therefore, we adopted these original
uncertainties, which include systematic error sources as described in
\citet{vf05}.  Based on this analysis, KELT-6 appears to be a main
sequence or very slightly evolved subgiant with $\teff = 6100
\pm 44$K, $\loggstar = 3.961 \pm 0.060$ and sub-solar metallicity, $\feh
= -0.277$. The star has a projected rotational velocity $\vsini = 5.0
\pm 0.5~{\rm km~s^{-1}}$.  

Comparing the parameter values determined from the TRES spectra using
SPC v2.2 to those determined from the HIRES spectra using SME, we
generally find agreement to $\sim 1\sigma$ or better, except for
$\vsini$, which differs by $\sim 3\sigma$. We do not have a good explanation for 
the $\vsini$ discrepancy. However, we do not use
$\vsini$ in our global fits, so this discrepancy is unimportant for
the present analysis.  The individual TRES spectra have a
signal-to-noise ratio (SNR) of $\sim40$ while the HIRES spectrum used
to derive the stellar parameters has a SNR of $\sim$180. We therefore
adopt the higher SNR HIRES stellar parameters for the analyses in this
paper, although we note that the uncertainties in both determinations
are likely to be dominated by systematic errors.

\subsection{UVW Space Motion\label{sec:motion}}

We evaluate the motion of KELT-6 through the Galaxy to place it among
standard stellar populations.  We adopt an absolute radial 
velocity of $+1.1 \pm 0.2$ km s$^{-1}$, based on the mean of the TRES and HIRES absolute RVs
listed in Table \ref{tab:rv1}, where the uncertainty is due to the systematic 
uncertainties in the absolute velocities of the RV standard stars. Combining the adopted absolute RV with
distance estimated from fitting the spectral energy distribution
(\S\ref{sec:sed}) and proper motion information from the NOMAD catalog
\citep{zacharias04}, we find that KELT-6 has $U,V,W$ space motion
(where positive $U$ is in the direction of the Galactic Center) of
$-6.3\pm 0.9$, $23.2\pm 0.8$, $6.9\pm 0.2$, all in units of km
s$^{-1}$, making it unambiguously a thin disk star.

\subsection{SED Analysis\label{sec:sed}}

\begin{figure}
\begin{center}
\includegraphics[scale=0.4]{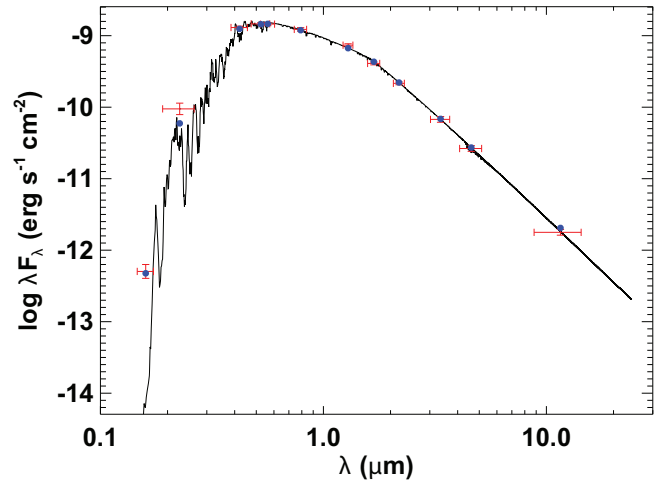}
\caption{Measured and best-fit SED for KELT-6 from UV through mid-IR.  The intersection points of the red error bars indicate measurements of the flux of KELT-6 in the UV, optical, NIR, and mid-IR passbands listed in Table \ref{tab:hostprops}. The vertical error bars are the $1\sigma$ photometric uncertainties, whereas the horizontal error bars are the effective widths of the passbands. The solid curve is the best-fit theoretical SED from the NextGen models of \citet{hau99}, assuming stellar parameters $\teff$, $\loggstar$ and $\feh$ fixed at the values in Table \ref{tab:fitpars} from the fiducial fit, with $A_{\rm V}$ and $d$ allowed to vary. The blue dots are the predicted passband-integrated fluxes of the best-fit theoretical SED corresponding to our observed photometric bands.\label{fig:sed}}
\end{center}
\end{figure}

We construct an empirical, broad-band spectral energy distribution (SED) of KELT-6,
shown in Figure \ref{fig:sed}.  We use the FUV and NUV fluxes from GALEX \citep{galex05}, the $B_{\rm T}$
and $V_{\rm T}$ colors from the Tycho-2 catalog \citep{hog00}, $V$ and $I_{\rm C}$ from TASS \citep{tass2000}, near-infrared
(NIR) fluxes in the $J$, $H$, and $K_{\rm S}$ passbands from the 2MASS Point Source
Catalog \citep{cutri03,skrutskie06}, and the near- and mid-IR fluxes in
three WISE passbands \citep{wright10}. We fit this SED to NextGen models from
\citet{hau99} by fixing the values of $\teff$, $\loggstar$ and $\feh$
inferred from the fiducial model fit to the light curve, RV, and
spectroscopic data as described in \S \ref{sec:exofast} and listed in
Table \ref{tab:fitpars}, and then
finding the values of the visual extinction $A_{\rm V}$ and distance $d$
that minimize $\chi^{2}$. The best fit model has a reduced $\chi^{2}$
of 1.61 for 10 degrees of freedom.  We find $A_{\rm V}$ = $0.01 \pm 0.02$ and $d$ = $222 \pm 8$ pc.
We note that the quoted statistical uncertainties on $A_{\rm V}$ and $d$ are
likely to be underestimated because we have not accounted for the
uncertainties in values of $\teff$, $\loggstar$ and $\feh$ used to
derive the model SED. Furthermore, it is likely that alternate model
atmospheres would predict somewhat different SEDs and thus values of
the extinction and distance.

\section{Characterization of the System}
\label{sec:exofast}

To determine the final orbital and physical parameters of the KELT-6
system, we combine the results from the spectroscopic analysis, the
light curves, and the HIRES RVs of KELT-6 as inputs to a global fit
using a custom version of EXOFAST \citep{eastman13}. The TRES RVs
are not used in the global fit analysis.
The EXOFAST analysis package does a
simultaneous Markov Chain Monte Carlo (MCMC) fit to the photometric
and spectroscopic data to derive system parameters.
It includes constraints on the stellar parameters $M_\star$ and
$R_\star$ from either the empirical relations in \citet{torres10} or from
Yonsei-Yale stellar models \citep{yy04}, in order to break the
well-known degeneracy between $M_\star$ and $R_\star$ for single-lined
spectroscopic eclipsing systems. EXOFAST scales the RV and light curve
data uncertainties such that the probability that the \chisq \ is
larger than the value we achieved, $\pchisq$, is $0.5$, to ensure the
resulting parameter uncertainties are roughly accurate. The global fit
method is similar to that described in detail in \citet{siv12}, but we
note a few differences below.\footnote{In the EXOFAST analysis, which includes the modeling of the filter-specific limb darkening parameters of the transit, we employ the transmission curves defined for the primed SDSS filters rather than the unprimed versions. We also use the Kepler transmission curve to approximate the CBB filter. We expect any differences due to those discrepancies to be well below the precision of all our observations in this paper and of the limb darkening tables from \citet{cb11}.}

\subsection{Light Curve Detrending\label{sec:detrend}}

Because KELT-6b's transits have an unusually long duration and
relatively shallow depth (by ground-based observing standards),
treatment of light curve systematics plays an important role in
the accuracy of parameters determined by the EXOFAST global fit. The
inclusion of detrending parameters into the global fit can often
mitigate the effect of light curve systematics, but sometimes at the
expense of introducing extra local minima in $\chi^{2}$ space, which
may cause other complications in the analysis.  Therefore, it is
important to maximize the detrending improvements to the fit of each
light curve while minimizing the number of detrending
parameters.  

Systematically fitting each light curve using all combinations of
$\sim$15 possible detrending parameters and comparing all of the
resulting $\chi^{2}$ values using the $\Delta\chi^{2}$ statistic would
be prohibitive. Instead, we opted to use the interactive
detrending capabilities of the AIJ package (see \S \ref{sec:phot}) to
search for up to three parameters that appeared to reduce the
systematics in each light curve. We then individually fit each of the
full transit light curves using EXOFAST, and repeated the fit using
various combinations of the detrending parameters selected for that
light curve. Finally, we compared $\chi^{2}$ from before and after the inclusion of an
additional detrending parameter to determine if the probability of 
a chance improvement was more than a few percent. If so, 
we did not include the additional detrending parameter in the global fit.

It is important to emphasize that the light curves fitted in EXOFAST
were the raw light curves (i.e. not the detrended light curves from
AIJ).  The only way in which the results of the AIJ analysis entered
into the final analysis was in the choice of detrending parameters and
the initial conditions adopted. Specifically, the detrend parameter
coefficients determined in AIJ were used as starting points for the
EXOFAST fits.  However these parameter coefficients were otherwise
allowed to vary freely in order to minimize $\chi^2$. 

One detrending parameter we included that warrants additional
discussion is an offset in the zero
point of the photometry arising from a change in placement of the
target and/or comparison star(s) on the CCD pixel array during time
series observations.  These positional changes typically result in a zero
point shift in the photometry at that epoch in the light curve due to
interpixel response differences and imperfect flat-field
corrections. We found such positional changes due to a meridian flip
in the MOCDK light curve on UT 2013-02-24, as well as an equipment
failure in the SPOT UT 2013-02-24 light curve (see Table
\ref{tab:photobs} and Figure \ref{fig:followup_lcs}).  We therefore
included a detrending parameter that accounts for a change in the zero
point of the relative photometry before and after the specified time.

In addition, fits to individual partial light curves often resulted in
obviously incorrect models.  We therefore chose detrending parameters
for such ingress- or egress-only data by hand using AIJ without a
rigorous $\Delta\chi^{2}$ analysis.

Light curves from near the time of predicted secondary eclipse were
treated somewhat differently.  In particular, these were airmass
detrended directly in AIJ, and when abrupt changes in the light curve
were correlated with a change in position of the target star on the
detector, $x$ and $y$ pixel positions of the target star centroid were
also used as detrending parameters.

The final detrending parameters adopted for all of the light curves are shown in
Table \ref{tab:photobs}.

\subsection{Global Fits\label{sec:global_fits}}

Using the KELT-6b primary transit light curves, the detrending
parameters and priors determined in the previous section, and the
results from the HIRES RV and spectroscopic analyses, we computed a
series of 12 global fits using our custom version of EXOFAST. The results of six
illustrative global fits are shown in Table \ref{tab:multipars}. The table lists
four global fit parameter choices (as detailed in the remainder of this subsection)
for each of the six fits, along with the values
of several key system parameters computed as part of each fit.

All global fits included a prior on orbital period $P=7.8457\pm0.0002$ days from
the KELT-North data and priors on host star effective temperature
$\teff=6100\pm44$~K and metallicity $\feh=-0.277\pm0.04$ from the
HIRES spectroscopy. The priors were implemented as a $\chi^{2}$ penalty
in EXOFAST (see \citealt{eastman13} for details).  For some of the global
fits we also included a prior on stellar surface gravity
$\loggstar=3.961\pm0.060$ from the HIRES spectroscopy. For the others, 
$\loggstar$ was constrained only by the transit data through the well-known
direct constraint on $\rho_\star$ from the light curve and RV data, combined with a constraint
on the stellar mass-radius relation through either the
Torres relations or the Yonsei-Yale evolutionary models. Fitting the
HIRES RV data independently to a Keplerian model, we found an acceleration (``RV slope'') of
$-0.239$ m~s$^{-1}$~day$^{-1}$, which is highly significant at the $\sim7\sigma$
level. Therefore, we proceeded with RV slope as a free parameter
for all global fits.

In addition to the slope, there were four additional choices that had to be
considered when performing the global fit. First, we needed to decide which transits to include in the
global fit.  We defined two alternative sets of light curve data to consider: (1)
the 5 ``full'' transits with both an ingress and egress and (2) all 16
full and partial transits. Second, as mentioned previously, we had the option to either
include a prior on stellar surface gravity $\loggstar=3.961\pm0.060$
based on the HIRES spectroscopy, or to fit for stellar surface gravity
without a prior. Third, we had the option to fit the orbital
eccentricity and argument of periastron as free parameters or fix them
to zero to force a circular orbit. Fourth, we had the option to break
the degeneracy between $M_\star$ and $R_\star$ by imposing external
constraints either from the relations of \citet{torres10} (Torres
constraints) or by imposing constraints from the Yonsei-Yale stellar
models \citep{yy04} (Yonsei-Yale constraints).

We first computed the four combinations of global fits using the 5
full transits with the Torres constraints. The four global fits are
defined by the different combinations of eccentric vs.\ circular
orbits, and $\loggstar$ with a spectroscopic prior vs.\ $\loggstar$ free. The column labeled 
``Fit 5'' in Table \ref{tab:multipars} shows the results for the Torres constrained, 
eccentric global fit, with no $\loggstar$ prior. As discussed in \S
\ref{sec:isochrones}, we plotted Yonsei-Yale stellar evolution tracks
corresponding to the stellar mass and metallicity results from these
global fits and found that the intersection of $\loggstar$ and
$\teff$ values from EXOFAST did not fall within $1\sigma$ of the
evolutionary tracks. We then computed the four combinations of global
fits using the 5 full transits with the Yonsei-Yale constraints and
found that for these fits the resulting $\loggstar$ and $\teff$
values were consistent with the corresponding Yonsei-Yale stellar
evolution tracks within $1\sigma$ error. Parameter values from these 
four fits are listed in the columns of Table \ref{tab:multipars} labeled 
``Fit 1'', ``Fit 2'', ``Fit 3'', and ``Fit 4''. The Torres constrained
planet mass and radius are larger than the Yonsei-Yale constrained mass
and radius by $\sim4-7\%$, and although we cannot determine if the
Torres relations or the Yonsei-Yale models best represent low
metallicity systems, we prefer the Yonsei-Yale constrained global fits for
self-consistency with the stellar evolution tracks in \S \ref{sec:isochrones}.

We next considered the 16 full and partial transit global fits. We
computed only the four combinations corresponding to the adopted
Yonsei-Yale constrained global fits. Although we computed very long Markov
chains with $10^6$ links, three of the four global fits resulted in
some parameters (mostly detrending parameters corresponding to partial
light curves) that did not fully converge. Converged parameters have
greater than 1000 independent draws and a Gelman-Rubin statistic less
than 1.01 (see \citealt{eastman13} and \citealt{ford2006}). The column labeled 
``Fit 6'' in Table \ref{tab:multipars} lists the results for the Yonsei-Yale constrained, 
eccentric global fit, with no $\loggstar$ prior. The system
parameters resulting from the 16 transit global fits are nearly
identical to the parameters from the 5 transit global fits. This is to be expected since detrended partial
light curves will not add significant constraints to transit depth and
shape when jointly fit with full transits. Given the partial transit
minor convergence issues, concerns about the ability
to properly remove systematics from these light curves, and the lack of significant additional
constraints on transit depth and shape from the partial transits, we
adopted the global fits based on the 5 full transits. We did however
use the 16 transit global fits for the transit timing analysis in
\S \ref{sec:ttvs}.

Next we examined the adopted Yonsei-Yale constrained global fits that
use only the 5 full transits. These four global fits are defined by
the different combinations of eccentric vs.\ circular orbits, and
$\loggstar$ prior vs.\ $\loggstar$ free. Since it is typically
difficult to measure $\loggstar$ to the same precision
spectroscopically that can be measured from a transit, we choose not
to impose a prior on $\loggstar$ from the HIRES spectroscopy. However,
we are wary of measurements of $\loggstar$ from the transits in this
case, since the duration is very long for a ground-based transit
observation. Comparing parameter values in column ``Fit 4'' of Table
\ref{tab:multipars} with column ``Fit 1'', and comparing column
``Fit 2'' with column ``Fit 3'', we found
that imposing a spectroscopic prior on $\loggstar$ increased the
stellar and planetary radii by $\sim3\%$ in the circular case and by
$\sim$7\% in the eccentric case. However, all of the system parameters
are within $\sim1\sigma$ of the results from the global fits without
a prior on $\loggstar$.

Since we had no strong prior expectation of tidal circularization of
KELT-6b's relatively long $\sim8$ day orbit, we adopted the more conservative eccentric
orbit global fits which have higher parameter errors. The eccentricity
resulting from a fit without a spectroscopic prior on $\loggstar$ is
$e=0.22_{-0.10}^{+0.12}$. The eccentricity resulting from a fit with
the HIRES spectroscopic prior on $\loggstar$ is
$e=0.27_{-0.12}^{+0.11}$. As pointed out by \citet{lsbias}, there is a
bias for inferred values of eccentricity with low significance, due to
the fact that $e$ is a positive definite quantity.  Although we adopt
an eccentric orbit global fit, we cannot exclude the hypothesis that
the orbit of KELT-6b is, in fact, circular.

Our final adopted fiducial stellar and planetary parameters were derived
from the 5 full transit, Yonsei-Yale constrained, eccentric orbit global fit with no
prior on $\loggstar$. Table \ref{tab:fitpars} lists the full set of system parameters for the fiducial fit.

Comparing the fiducial system parameters with those from the other 11 global fits,
we note differences in 
planetary mass $\Delta M_P\sim10$\% ($\sim1\sigma$), 
planetary radius $\Delta R_P\sim10$\% ($\sim1\sigma$),
orbital radius $\Delta a\sim5$\% ($\sim4\sigma$),
planetary equivalent temperature $\Delta$$T_{\rm{eq}}\sim5$\% ($\sim1\sigma$),
stellar mass $\Delta M_\star\sim15$\% ($\sim3\sigma$), 
and stellar radius $\Delta R_\star\sim15$\% ($\sim1.5\sigma$).
Clearly, the choice of global fit input parameters, priors, and 
external constraints, significantly affects some of the inferred 
system parameters. Thus, it is important to note that other plausible global fits yield  
significantly different values for some system parameters.

The HIRES RV uncertainty scaling for the fiducial global fit is 2.808,
which is fairly high and is suggestive of substantial stellar jitter in the RV
data. The RMS of the RV residuals of the fit to these scaled data is
8.0 m s$^{-1}$, which is somewhat high ($\sim2\sigma$) compared to what we would 
expect based on \citet{wright05}. We do not have a compelling explanation for the high RV residuals. 
As noted in \S \ref{sec:RV}, we did not attempt to
measure line bisectors for the HIRES data.

\subsection{Transit Timing Variations\label{sec:ttvs}}

We investigated the transit center times of the 16 full and partial
transits adopted from the 16 transit, Yonsei-Yale constrained,
eccentric orbit global fit with no prior on $\loggstar$ for any signs
of transit time variations (TTVs). We were careful to ensure that
all quoted times had been properly reported in $\bjdtdb$ (e.g.,
\citealt{east10}). When we performed the global fit, we allowed for
transit time $T_{\rm C,i}$ for each of the transits shown in Table \ref{tab:ttvs}
to be a free parameter.  Therefore, the individual follow-up transit light curves
do not constrain the KELT-6b ephemeris (global epoch $T_{\rm C}$ and period $P$).
Rather, the constraints on these parameters in the global fit come only from the RV
data, and the prior imposed from the KELT discovery data.
Using the follow-up transit light curves to constrain the ephemeris
in the global fit would artificially reduce any observed TTV signal.

\begin{figure}
\begin{center}
\includegraphics[scale=1.15]{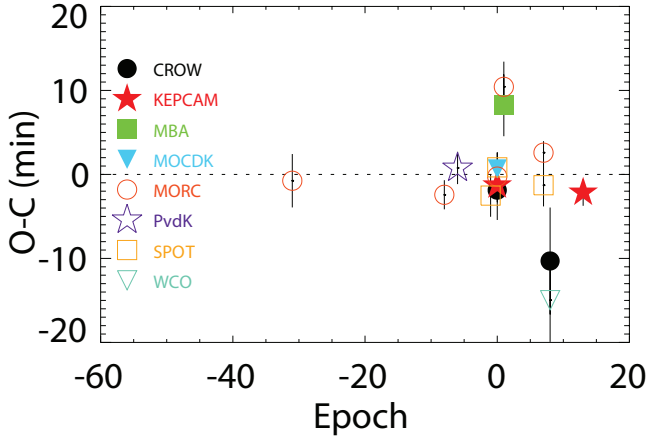}
\caption{The residuals of the transit times from the best-fit ephemeris. The transit times are given in Table \ref{tab:ttvs}.
The observatory/telescope abbreviations are the same as in Table \ref{tab:photobs}.
}\label{fig:ttvs}
\end{center}
\end{figure}

Subsequent to the global fit, we then derived a separate ephemeris from
only the transit timing data by fitting a straight line to all
inferred transit center times from the global fit.  These times are
listed in Table \ref{tab:ttvs} and plotted in Figure \ref{fig:ttvs}.
We find $T_{\rm 0} = 2456347.796793 \pm 0.000364$, $P_{\rm Transit} = 7.8456314 \pm
0.0000459$, with a $\chi^2$ of 38.70 and 14 degrees of freedom. While
the $\chi^2$ is larger than one might expect, this is often the case
in ground-based TTV studies, likely due to systematics in the transit data. There
are $\sim3\sigma$ deviations from the linear ephemeris on epochs 1
and 8. However, although there are consistent TTV measurements from
two independent observatories on both of those epochs, we note that
these data are all from ingress or egress only observations.  Given
the likely difficulty with properly removing systematics in partial
transit data, we are unwilling to claim convincing evidence for TTVs.
Further study of KELT-6b transit timing is required to rule out TTVs.

\begin{deluxetable}{lccrrc}
\tabletypesize{\scriptsize}
\tablecaption{KELT-6 Transit Times\label{tab:ttvs}}
\tablewidth{0pt}
\tablehead{\colhead{Epoch} & \colhead{$T_{\rm C}$} & \colhead{$\sigma_{\rm T_{\rm C}}$} & \colhead{O-C} & \colhead{O-C} & \colhead{Observatory/}\\
\colhead{~} & \colhead{\bjdtdb} & \colhead{Sec} & \colhead{Sec} & \colhead{$\sigma_{\rm T_{\rm C}}$} & \colhead{Telescope}}
\startdata
 -31 &  2456104.581654  &  190  &   -45.17  &  -0.24  &  MORC\\
  -8 &  2456285.030078  &  104  &  -145.43 &  -1.40  &  MORC\\
  -6 &  2456300.723507  &  115  &   46.31 &   0.40  &  PvdKO\\
  -1 &  2456339.949428  &  151  &  -151.18 &  -1.00  &  SPOT\\
   0 &  2456347.797243  &  118   &  39.38 &   0.33  &  MOCDK\\
   0 &  2456347.796609  &   79   &  -15.31 &  -0.19 &   MORC\\
   0 &  2456347.797368  &  104  &   50.44  &  0.48  & SPOT\\
   0 &  2456347.795916  &  120  &  -77.26 &  -0.64  &  KEPCAM\\
   0 &  2456347.795513  &  213  &  -112.42 &  -0.53  &  CROW\\
   1 &  2456355.649675  &  179  &  625.66 &   3.48  &  MORC\\
   1 &  2456355.648134  &  220  &  493.04 &   2.24  &  MBA\\
   7 &  2456402.718026  &   82 &   155.28  &  1.89  &   MORC\\
   7 &  2456402.715318  &  151  &  -77.39 &  -0.51   &  SPOT\\
   8 &  2456410.554740  &  381  &  -618.29  & -1.62   &  CROW\\
   8 &  2456410.550639  &  298  &  -898.14  & -3.01  &   WCO\\
 13 &  2456449.788514  &   95  &   -128.47  & -1.35  &  KEPCAM\\
\enddata
\tablecomments{
The observatory/telescope abbreviations are the same as in Table \ref{tab:photobs}.
}
\end{deluxetable}

\section{Evidence for a Tertiary Companion}
\label{sec:tertiary}
\begin{figure}
\begin{center}
\includegraphics[scale=1.03]{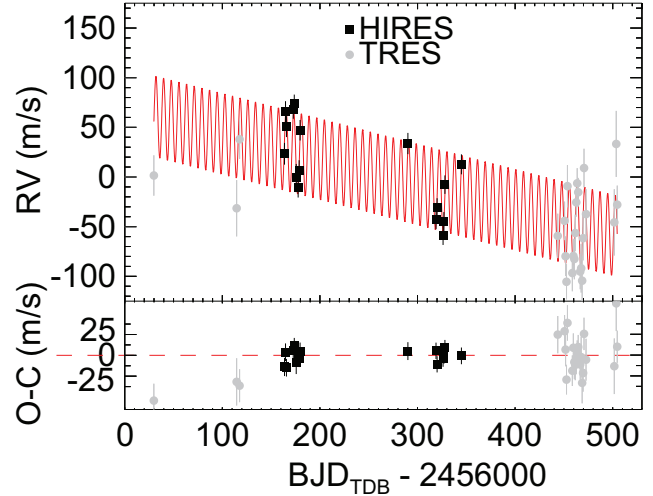}
\caption{HIRES and TRES unphased KELT-6 radial velocities.  HIRES radial velocity measurements are shown as black squares. TRES radial velocities are show as gray circles. The HIRES error has been scaled by 2.808 as determined by the fiducial EXO\-FAST global fit (see \S \ref{sec:exofast}). The TRES errors are unrescaled. The single-planet plus linear slope fiducial model of the KELT-6 system fit to the HIRES data only is shown as a solid red line. The TRES RVs have been shifted by a constant offset that best-fits the fiducial model. Although the TRES data appear to indicate a turnover in the RV slope, a joint fit to the HIRES and TRES data indicate only
marginal evidence for a turn-over (see \S \ref{sec:tertiary}). \label{fig:RV_unphased}}
\end{center}
\end{figure}

The Keck HIRES radial velocities show a downward trend that is well
modeled by a linear slope over the time span of the HIRES RVs as illustrated
in Figure \ref{fig:RV_unphased}. The fiducial model, which is displayed 
as a solid red line, is fit to the HIRES data only and has a slope of
$\dot{\gamma}=-0.239\pm0.037~{\rm m~s^{-1}~day^{-1}}$. A two-planet
fit with the tertiary in a circular orbit yields a negligible
improvement of $\Delta\chi^2 = 2.2$ relative to the fit with constant
acceleration, which has a $\sim30\%$
probability of happening by chance. With the inclusion of the full set
of 24 re-reduced TRES RVs (see \S \ref{sec:RV}) into the single-planet
plus slope and two-planet fits, $\Delta\chi^2 = 3.8$, which has a
$\sim15\%$ probability of happening by chance. Although the TRES RVs
shown in Figure \ref{fig:RV_unphased} appear to fairly strongly 
indicate a turn-over in the RV slope, the statistical analysis above finds only
marginal evidence for a turn-over. The TRES RVs shown in Figure \ref{fig:RV_unphased}
have been shifted to best fit the HIRES fiducial model. Characterization of the 
tertiary will require continued RV monitoring of the KELT-6 system.

\begin{figure}
\begin{center}
\includegraphics[scale=0.33]{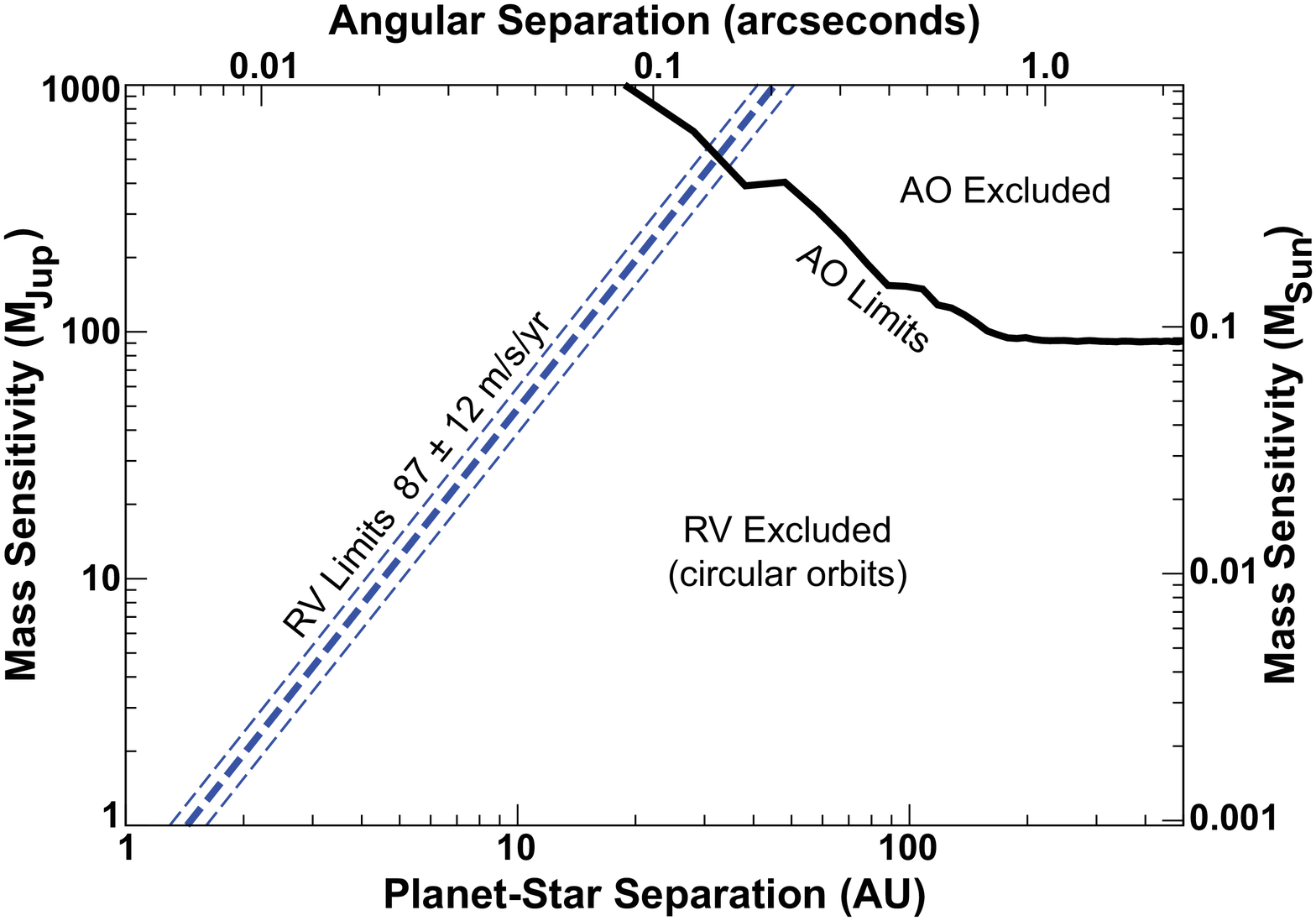}
\caption{KELT-6 tertiary mass limits derived from the Keck AO image and 
measured HIRES projected acceleration versus separation in AU. The
top scale shows angular separation in arcseconds corresponding
to a given projected separation, assuming a distance of $222$pc to 
the system. The AO mass limits as a function
of projected separation are shown by the heavy solid black line. For a given projected
separation, masses above the heavy solid black line are
excluded. The heavy blue dashed line shows the lower limit on mass
of the tertiary that could cause the observed projected
acceleration, as a function of semimajor axis, and
assuming circular orbits.  The light blue dashed lines show the $1\sigma$
uncertainty in the limit due to the uncertainty in the projected
acceleration. Assuming that the systematic radial velocity has varied
monotonically between the two groups of HIRES RVs shown in Figure
\ref{fig:RV_unphased}, masses for the tertiary causing the acceleration that fall
below the dashed blue lines are excluded. However, there could be
undetected companions in the region below the blue dashed lines that
are not responsible for the observed RV
acceleration.\label{fig:mass}}
\end{center}
\end{figure}

Our Keck AO $K'$ image shows no significant detection of off-axis
sources, although there are a couple of speckles at the threshold of
detection (see Figure \ref{fig:ao} and \S \ref{sec:ao}). Figure
\ref{fig:mass} shows the limits on mass from the AO image and from the
HIRES RVs. For a given projected separation, masses above the heavy
solid black line are excluded by the AO image. The heavy blue dashed
line shows the lower limit for mass of the tertiary for circular
orbits as a function of semimajor axis implied by the projected
acceleration of ${\cal A} = 87\pm12~{\rm m~s^{-1}~yr^{-1}}$ measured
from the HIRES RV data.  For a circular orbit with semimajor axis $a$
and a given minimum planet mass $M_{\rm P} \sin i$, the maximum projected
acceleration of the star due to the planet occurs at conjunction (or opposition), and is
${\cal A}=G M_{\rm P} \sin i~a^{-2}$ \citep{torres99}.  Thus a strict lower
limit on the tertiary mass capable of producing the measured
acceleration can be defined for a given $a$, assuming circular
orbits\footnote{We note that this constraint assumes that the tertiary
imposes a constant acceleration during the time spanned
by the RV observations.  In particular, it assumes that the systemic radial velocity
has varied monotonically between the two groups of HIRES RVs shown in Figure
\ref{fig:RV_unphased}. Because there is a substantial gap between these two groups of points,
shorter-period orbits for the tertiary in which the acceleration changes sign twice between
the two groups are possible.  However, we deem these to be unlikely.}. 
Note that this mass increases as the square of projected separation. 
The light blue dashed lines show the $1\sigma$
uncertainty on the minimum $M_{\rm P} \sin i$ due to the uncertainty in the measured acceleration. 
Masses for the purported tertiary that fall below the blue dashed lines are
excluded, as they do not provide sufficient acceleration at conjunction for a given
semimajor axis to explain the observed trend even for an edge-on orbit.
However, there could be undetected companions in the region below the blue dashed lines that are not  
responsible for the observed RV acceleration.
The RV and AO mass curves intersect for masses comparable to
the primary star, and at the diffraction-limit of a 10 m telescope
on the projected separation axis. Therefore, if the speckles at the
threshold of detection in Figure \ref{fig:ao} are astrophysical, they
cannot be responsible for the observed long-term acceleration in the
KELT-6 radial velocities.

\vspace{0.25in}

\section{False Positive Analysis}
\label{sec:falsepos}

One of the many challenges of ground-based photometric surveys for transiting
planets is the relatively high rate of astrophysical false positives prior to RV 
and high precision photometry follow-up observations (e.g. \citealt{latham2009}).
Blended eclipsing stellar binary or triple systems can mimic some
of the observable signatures of transiting low-mass companions to
single stars.  \citet{brown2003} estimated the a priori detection rates of such false positives
in ground-based transit surveys similar to KELT, finding a rate that was a factor of several
times larger than the expected detection rate for transiting giant planets.  However,
for KELT-6b, we have several lines of evidence that disfavor a false positive scenario.  

First, we measured the line bisector spans of the TRES spectra following
\citet{torres2007} to explore the possibility that the RV variations are 
actually distortions in the spectral line profiles due to a nearby unresolved
eclipsing binary or stellar activity. The bisector span variations are listed in Table \ref{tab:rv1} 
and plotted in the bottom panel of Figure \ref{fig:rv}. The resulting bisector span variations
are consistent with zero and show no correlation with the RV variations. As noted in \S \ref{sec:RV}, 
we did not attempt to measure line bisectors for the HIRES spectra since
the PSF varies quite dramatically in the slit-fed HIRES instrument
simply from guiding and spectrometer focus variations, which can cause instrumentally induced line
asymmetries that cannot be easily distinguished from stellar sources.

Second, our follow-up photometric observations of full transits in
several different filters ($griz$) are all consistent with the primary
transit having nearly the same depth, and are well-modeled by transits
of a dark companion across a star with the limb darkening consistent
with its spectroscopically measured $\teff$ and $\loggstar$ (see
Figure \ref{fig:followup_lcs} and \S \ref{sec:global_fits}). Since the 
multi-band depth difference expected for a false positive scenario 
depends strongly on the color difference of the blended stars, 
the multi-band transit observations cannot rule out all false positive 
configurations, but can significantly limit the allowed parameter space.

\begin{figure}
\begin{center}
\includegraphics[scale=0.69]{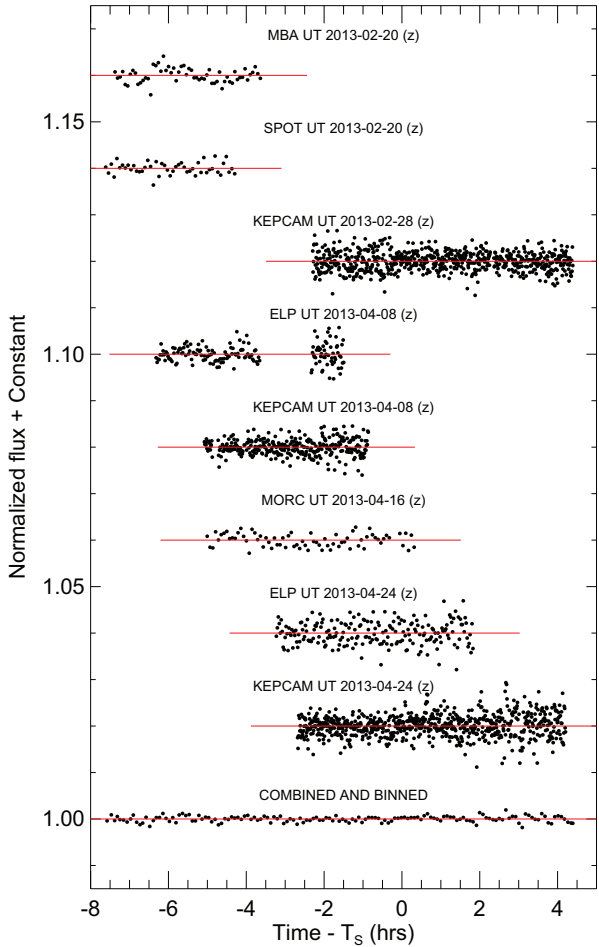}
\caption{Phased observations of KELT-6 near the time of predicted secondary transit.  The ephemeris used to phase the data is $T_0=2456265.51$ $(BJD_{TDB})$ and $P=7.8457$ (days).  The fiducial ephemeris is uncertain by $\sim0.7$ days. Our observations cover only $\sim50\%$ of the region of uncertainty. The red overplotted lines are the constant brightness models. The observatory/telescope abbreviations are the same as in Table \ref{tab:photobs}. The bottom light curve shows all observations combined and binned in 5 minute intervals, and has residuals of 0.06\% RMS. We find no evidence for a secondary transit in the data. \label{fig:secondary}}
\end{center}
\end{figure}

Third, we collected eight sequences of photometric observations near
the time of predicted secondary eclipse (at five different epochs) in
$z$ and Pan-STARRS-Z bands as detailed in Table \ref{tab:photobs}.
The individual phased light curves and the combined binned light curve
are shown in Figure \ref{fig:secondary} and cover 12 hours near the
time of predicted secondary eclipse. As shown in Table
\ref{tab:fitpars}, the fiducial predicted time of secondary eclipse
has an uncertainty of $\sim16$ hours. We do not find conclusive
evidence of a $\ga 1$ mmag secondary eclipse ingress or egress in our
data. However, we do not have complete phase coverage of all the 
secondary eclipse times that are allowed by our global fits, and 
therefore we cannot place a robust lower limit on the depth of 
any putative secondary transit arising from a blended eclipsing binary. 

Although the multi-band transit and secondary eclipse observations cannot
exclude all blend scenarios, they disfavor blend scenarios
in which the observed transits are due to diluted eclipses of a much
fainter and redder eclipsing binary (e.g., \citealt{odonovan2006}).

Fourth, the fiducial transit derived stellar surface gravity $\log{g_{\star_{\rm transit}}}=4.074_{-0.070}^{+0.045}$ (the fiducial fit does not use a spectroscopic prior on $\loggstar$) and the HIRES spectroscopically derived surface gravity $\log{g_{\star_{\rm HIRES}}}=3.961\pm0.060$ are consistent within $\sim1.5\sigma$.

Finally, our adaptive optics imaging excludes companions beyond a distance of 0.5 arcseconds from KELT-6 down to a magnitude difference of 6.0 magnitudes at $10\sigma$ confidence. See Figure \ref{fig:contrast}.

We conclude that all of the available data are best explained by a Jupiter-sized, Saturn-mass companion transiting a slowly-rotating late-F star, with little or no evidence for significant contamination from blended sources.

\section{Evolutionary Analysis}
\label{sec:evolv}

\subsection{Stellar Models and Age}
\label{sec:isochrones}
\begin{figure}
\begin{center}
\includegraphics[scale=0.55]{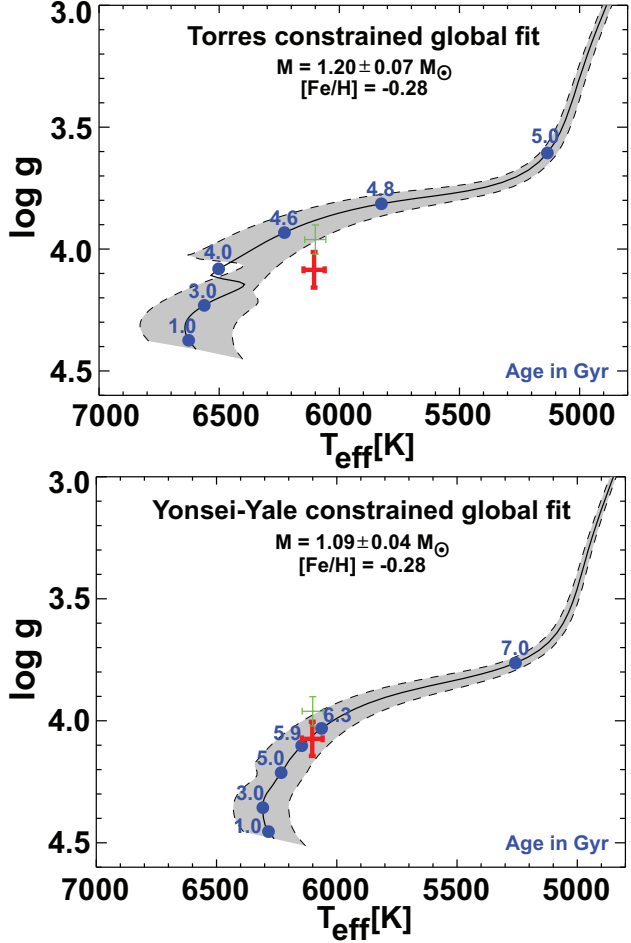}
\caption{Theoretical HR diagrams based on Yonsei-Yale stellar evolution models \citep{yy04}. The gray swaths represent the evolutionary track for the best-fit values of the mass and metallicity of the host star from the global fits corresponding to Table \ref{tab:multipars} columns ``Fit 1'' (bottom panel) and ``Fit 5'' (top panel) and discussed in \S \ref{sec:exofast}. The tracks for the extreme range of $1\sigma$ uncertainties on $M_\star$ and $\feh$ are shown as dashed lines bracketing each gray swath. {\it Top panel:} The Yonsei-Yale track based on the Torres constrained global fit corresponding to Table \ref{tab:multipars} column ``Fit 5' (see \S \ref{sec:evolv} for explanation). {\it Bottom panel:} The Yonsei-Yale track based on a Yonsei-Yale constrained fiducial global fit corresponding to Table \ref{tab:multipars} column ``Fit 1'. The thick red crosses show $\teff$ and $\loggstar$ from the EXOFAST global fit analyses. The thin green crosses show the inferred $\teff$ and $\loggstar$ from the HIRES spectroscopic analysis alone. The blue dots represent the location of the star for various ages in Gyr. The Torres constrained global fit is inconsistent with the Yonsei-Yale track at $>1\sigma$. We adopt the Yonsei-Yale constrained global fit represented in the bottom panel resulting in a slightly evolved star with an estimated age of $6.1 \pm 0.2$ Gyr, where the uncertainty does not include possible systematic errors in the adopted evolutionary tracks. \label{fig:hrd}}
\end{center}
\end{figure}

We use global fit values for $\teff$, $\loggstar$, stellar mass, and metallicity (\S \ref{sec:exofast} and Table \ref{tab:multipars} columns ``Fit 1'' and ``Fit 5''), in combination with the theoretical evolutionary tracks of the Yonsei-Yale stellar models \citep{yy04}, to estimate the age of the KELT-6 system. We have not directly applied a prior on the age, but rather have assumed uniform priors on [Fe/H], $\loggstar$, and $\teff$, which translates into non-uniform priors on the age. The standard version of EXOFAST uses the \citet{torres10} relations to estimate stellar mass and radius at each step of the MCMC chains. The top panel of Figure \ref{fig:hrd} shows the theoretical HR diagram ($\loggstar$ vs.\ $\teff$) corresponding to Table \ref{tab:multipars} column ``Fit 5''. We also show evolutionary tracks for masses corresponding to the $\pm1\sigma$ extrema in the estimated uncertainty. The Torres constrained global fit values for $\teff$ and $\loggstar$ are inconsistent by more than $1\sigma$ with the Yonsei-Yale track corresponding to the stellar mass and metallicity preferred by this global fit. To investigate the inconsistency, we modified EXOFAST to use the Yonsei-Yale models rather than the \citet{torres10} relations to estimate stellar mass and radius at each MCMC step. The bottom panel of Figure \ref{fig:hrd} is the same as the top panel, but for the fiducial Yonsei-Yale constrained global fit corresponding to Table \ref{tab:multipars} column ``Fit 1''. The intersection of global fit values for $\teff$ and $\loggstar$ now fall near the Yonsei-Yale track at $6.1 \pm 0.2$ Gyr, where the uncertainty does not include possible systematic errors in the adopted evolutionary tracks. The Torres constrained global fit yields an age that is about 25\% younger, and planet mass and radius that is larger by $\sim4-7\%$.  Although we cannot explain the inconsistency between the Torres constrained global fit and the Yonsei-Yale track, we expect that it may be due to slight inaccuracies in the Yonsei-Yale models and/or the \citet{torres10} relations for metal poor stars. We adopt the Yonsei-Yale constrained global fit for the analyses in this paper.

KELT-6 is evidently a late-F star that is just entering the subgaint stage of evolution. To check that the isochrone age is consistent with other parameters of KELT-6, we use the gyrochronology relations of \citet{barnes07} to compute the age based on the rotation period of the star and its $B-V$ color. We checked the KELT light curve for periodic variability associated with spot modulation as an indicator of $P_{rot}$, but we were unable to detect any significant sinusoidal variability beyond the photometric noise. Lacking a direct measurement, we estimated $P_{rot}$ using the projected rotational velocity from \S \ref{sec:spec_params} and the stellar radius from the adopted global fit in \S \ref{sec:exofast} to be $P_{\rm rot}/\sin i_{\rm rot} = 16.2 \pm 3.8$ days. \citet{harris64} photoelectrically measured magnitudes and colors of KELT-6 and found $B-V=0.49\pm0.008$. Tycho \citep{hog00} measured $B_{\rm T}$ and $V_{\rm T}$ (Table \ref{tab:hostprops}), and through the filter transformations described in \citet{esa97}, the Tycho-based color is $B-V=0.415\pm0.069$. Because the \citet{harris64} precision is much higher than Tycho's, and since the Tycho color is consistent with the \citet{harris64} color at nearly $1\sigma$, we adopt the \citet{harris64} color for this analysis.  In particular, we
are worried about inaccuracies in the Tycho-to-Johnson filter-band
transformations, especially for metal-poor stars; \citet{hog00} state
that these filter-band transformations are approximate.  Based on the
adopted rotation period and $B-V$ color of the star, we calculate the
maximum predicted age (subject to the inclination of the rotation axis
to our line of sight) to be $5.7 \pm 1.3$ Gyr, which is fully
consistent with the isochrone age.  We note that if the Tycho fiducial
color is used with the adopted rotation period, the \citet{barnes07}
relations yield an unrealistically large age of 46 Gyr, due to the
fact that these relations break down for stars with $B-V \la 0.4$,
which generally have small or non-existant convective envelopes.

\subsection{Insolation Evolution}
\label{sec:orbevol}

In an investigation of transiting giant exoplanets, \citet{ds11} found that for planets insolated beyond the  threshold of $2 \times 10^8$ erg s$^{-1}$ cm$^{-2}$ the radii are inflated compared to those planets with lower levels of insolation.  KELT-6b currently has incident flux well above that threshold, and is a mildly inflated hot Saturn with a density of $0.248_{-0.050}^{+0.059}$ g cm$^{-3}$. It follows the insolation-inflation trend displayed in Figure 1 of \citet{ds11}. However, it is worth investigating whether KELT-6b has always been insolated above the \citet{ds11} threshold.  If it turns out that KELT-6b only recently began receiving enhanced irradiation, this could provide an empirical probe of the timescale of inflation mechanisms (see \citealt{assef09} and \citealt{sm12}).

To answer that question, we simulate the reverse and forward evolution of the star-planet system, using the fiducial global fit parameters listed in Table \ref{tab:fitpars} as the present boundary conditions. This analysis is not intended to examine circularization of the planet's orbit, tidal locking to the star, or any type of planet-planet or planet-disk interaction or migration. Rather, it is a way to infer the insolation of the planet over time due to the changing luminosity of the star and changing star-planet separation.

\begin{figure}
\begin{center}
\includegraphics[scale=0.4,trim=0mm 0mm 0mm 0mm,clip,angle=0]{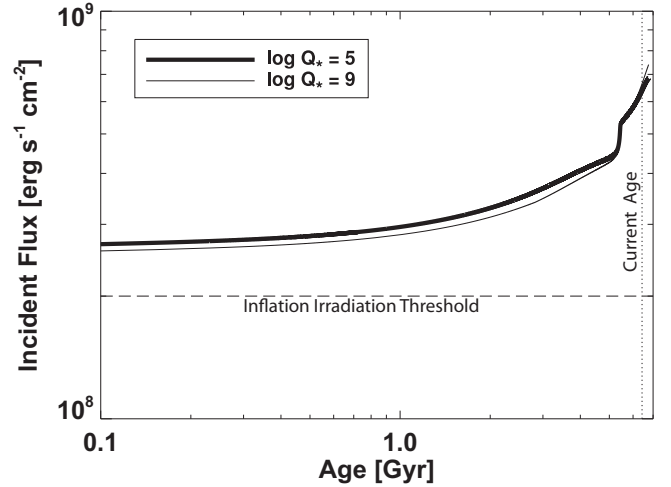}
\caption{Change in incident flux for KELT-6b, with test values of log$~Q_{\star}=5$ and log$~Q_{\star}=9$  for KELT-6.  This system is clearly insensitive to the value of $Q_{\star}$ in the range we analyzed. In both cases, the planet has always received more than enough flux from its host to keep the planet irradiated beyond the insolation threshold of $2 \times 10^8$ erg s$^{-1}$ cm$^{-2}$ identified by \citet{ds11}.\label{fig:evolv}}
\end{center}
\end{figure}

We include the evolution of the star, which is assumed to follow the
YREC stellar model corresponding to $M=1.1$ $M_{\odot}$ and $Z=0.0162$
\citep{siess00}.  We also assume that the stellar rotation was
influenced only by tidal torques due to the planet, with no magnetic
wind and treating the star like a solid body. Although the fiducial model from \S \ref{sec:global_fits}
has an eccentric orbit, we assume a circular
orbit throughout the full insolation analysis.  The results of our simulations
are shown in Figure \ref{fig:evolv}.  We tested a range of values for
the tidal quality factor of the star $Q_{\star}$, from
log$~Q_{\star}$ = 5 to log$~Q_{\star}$ = 9.  We find that this
system is highly insensitive to the value of $Q_{\star}$,
because tides are not important for this system for the parameter
ranges we analyzed. In all cases, KELT-6b has always received more than
enough flux from its host to keep the planet irradiated beyond the \citet{ds11}
insolation threshold required for inflation.

\section{Discussion}
\label{sec:discuss}
From our global fit to the spectroscopy, light curves, and HIRES RVs, we find that KELT-6b
is a metal-poor hot Saturn with a measured mass
$M_{\rm P}=0.430_{-0.046}^{+0.045}~\mjup$ and radius
$R_{\rm P}=1.193_{-0.077}^{+0.130}~\rjup$.  It is on an orbit with eccentricity $e=0.22_{-0.10}^{+0.12}$ and
semimajor axis of $a=0.07939_{-0.00099}^{+0.00100}$~AU.  The host KELT-6 is a slightly evolved
late-F star with a mass $M_\star=1.085\pm0.043~\msun$, radius
$R_\star=1.580_{-0.094}^{+0.16}~\rsun$, effective temperature
$\teff=6102\pm 43~{\rm K}$, and a likely age of $6.1\pm0.2$~Gyr.  
Because of its larger semimajor axis (compared to a typical hot Jupiter), KELT-6b receives a moderate stellar insolation
flux of $\langle F \rangle = 6.53_{-0.76}^{+0.92} \times 10^8~{\rm
erg~s^{-1}~cm^{-2}}$, implying a moderate equilibrium temperature of $T_{\rm eq}=1313_{-38}^{+59}~{\rm
K}$ assuming zero albedo and perfect redistribution.  The surface gravity and density of KELT-6b are
$\loggp=2.868_{-0.081}^{+0.063}$ and $\rho_{\rm p}=0.311_{-0.076}^{+0.069}~{\rm g}~{\rm cm}^{-3}$.  We do not have
in-transit KELT-6b RV data, so we have no Rossiter-McLaughlin effect constraint on the projected rotation
axis of its host star.

\begin{figure}[!htbp]
\begin{center}
\includegraphics[scale=0.30,angle=0]{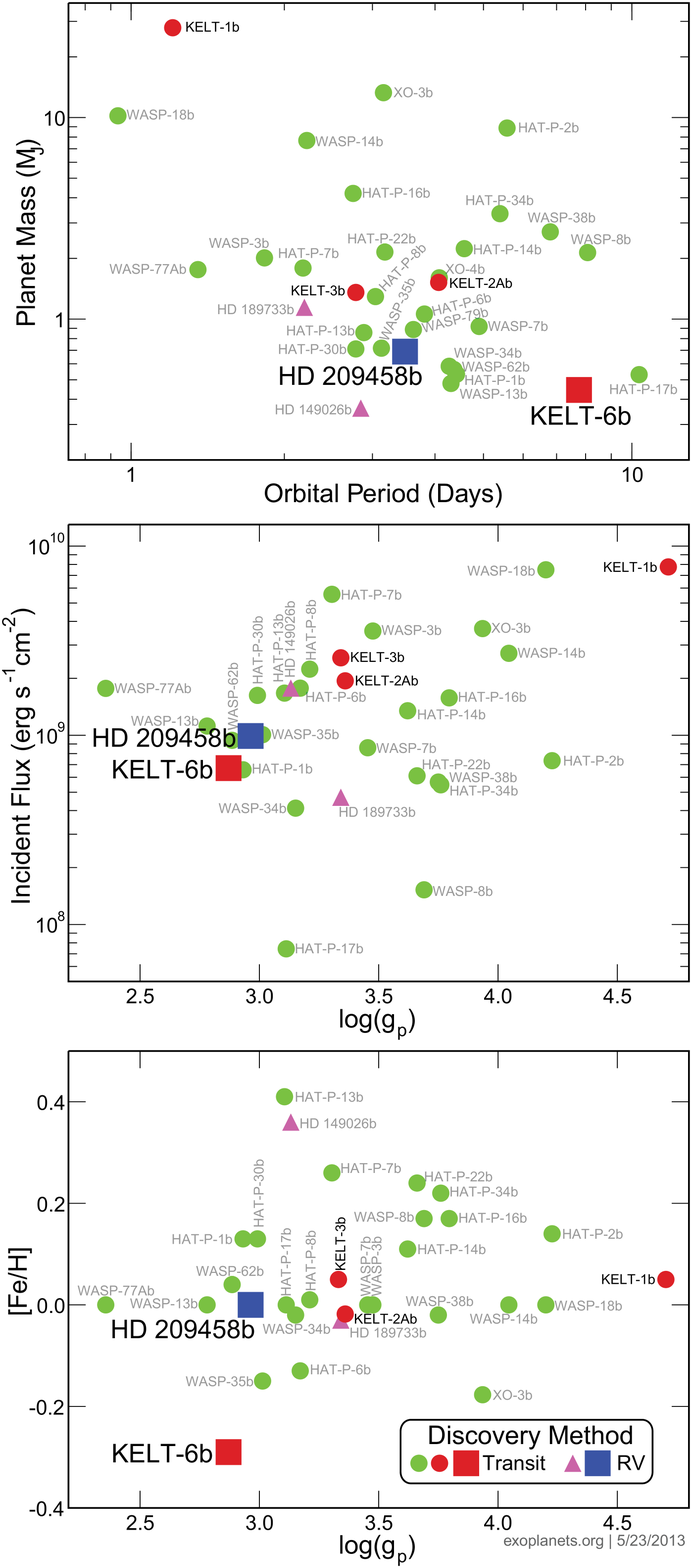}
\caption{Comparisons of bright, transiting, hot gas giants with $m>0.1~\mjup$, $P<20$ days, and host star $V<11.0$. The three RV discovered planets are shown as magenta filled triangles (HD~189733b and HD~149026b) and a large blue filled square (HD~209458b). All other transits were discovered by ground-based transit surveys. No Kepler targets currently meet the specified criteria for inclusion in the group. The KELT-north survey planets are shown as red filled circles, except KELT-6b which is shown as a large red filled square. All other planets are shown as green filled circles. {\it Top panel:} Planet mass as a function of the orbital period. Both KELT-6b and HD~189733b are sub-Jupiter mass planets.  {\it Middle panel:} Incident flux as a function of planet surface gravity. KELT-6b has surface gravity and incident flux similar to HD~209458b. All else being equal, objects in the top left have the highest transmission spectroscopy signal. {\it Bottom panel:} $\feh$ as a function of planet surface gravity. KELT-6b has metallicity lower than HD~209458b by $\sim0.3$ dex. KELT-6b and HD~209458b offer an opportunity to perform a comparative measurement of two similar planets in similar environments around stars of very different metallicities.\label{fig:comps}}
\end{center}
\end{figure}

Even among the ever growing list of known transiting exoplanets,
KELT-6b is unique. In Figure \ref{fig:comps} we compare planet mass as
a function of the orbital period (top panel), incident flux as a
function of $\loggp$ (middle panel), and $\feh$
as a function of $\loggp$ (bottom panel), for
the group of all transiting hot gas giants orbiting bright hosts,
which we define as $m>0.1~\mjup$, $P<20$ days, and host star
$V<11.0$. Within that group, KELT-6 is among the 20 brightest host
stars, and KELT-6b has the third longest orbital period (top panel),
second lowest mass (top panel), and is the most metal-poor (bottom
panel). In the larger group of all transiting exoplanets discovered by 
ground-based transit surveys, KELT-6b has the sixth longest period 
and the second longest transit duration. To our knowledge, the
high precision photometric follow-up observations reported in this
work include the longest duration transit ever fully observed from a
single ground-based telescope.

\begin{figure}[!htbp]
\begin{center}
\includegraphics[scale=0.44,angle=0]{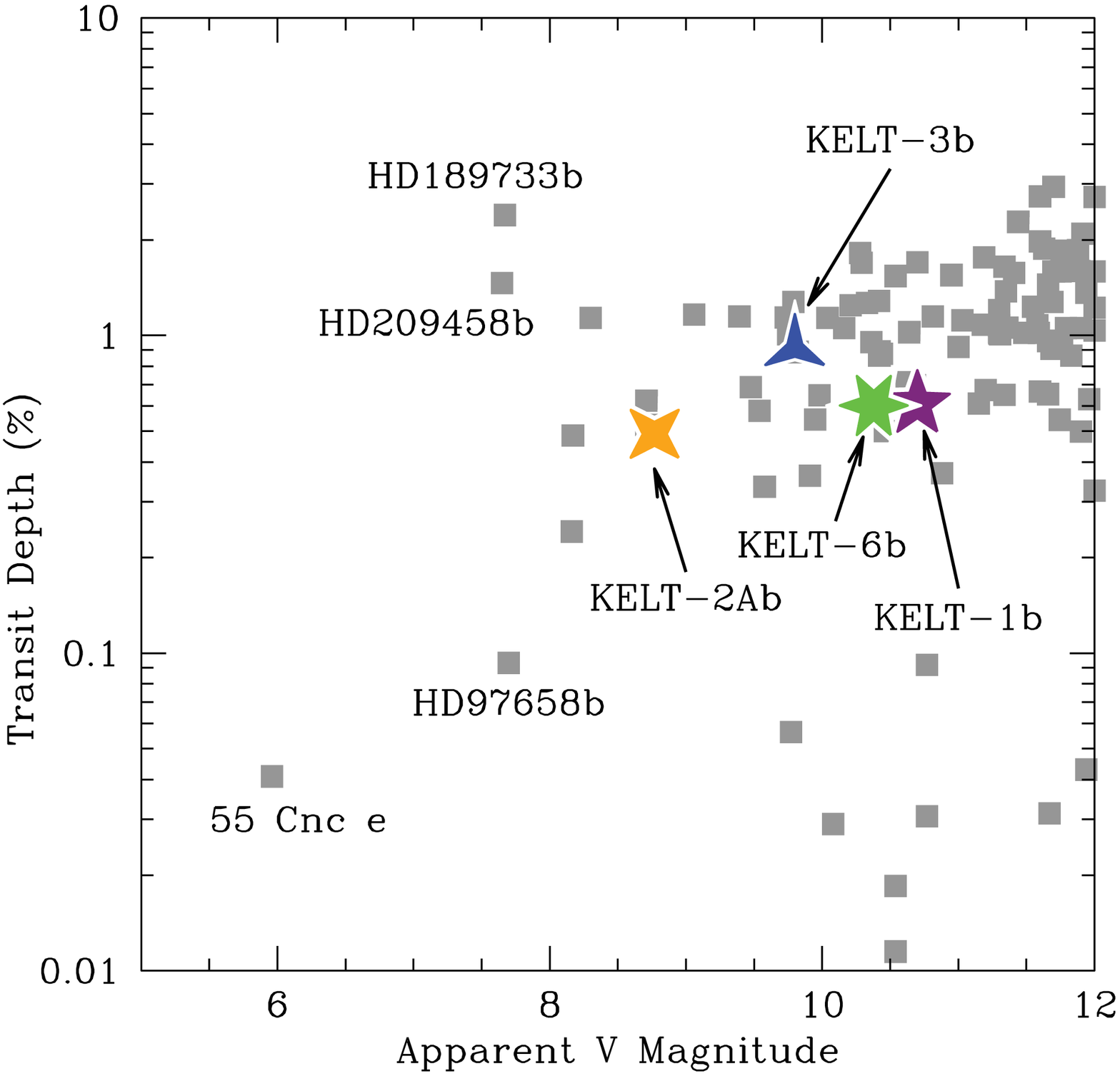}
\caption{Transit depth assuming no limb darkening as a function of the
host star apparent $V$ magnitude for transiting systems with relatively
bright ($V\le 12$) hosts.  KELT-6b is shown as the green six-pointed
star.  The other KELT discoveries are also shown, and the transiting
systems with very bright hosts ($V\le 8$) are labeled. Systems in the top
left tend to be the most amenable to detailed spectroscopic and photometric studies.
\label{fig:rprsv}}
\end{center}
\end{figure}

Perhaps the most significant importance of the KELT-6b discovery is
that it has similar $\loggp$ and incident flux as HD~209458b (middle
panel), one of the most studied and best understood exoplanets, but
its host has a metallicity that is lower than HD~209458 by $\sim0.3$
dex\footnote{While HAT-P-1b, WASP-13b, WASP-35b, and WASP-62b 
have $\loggp$ and incident flux similar to HD~209458b, none of them 
are metal poor except for WASP-35b, which has a metallicity of $\feh=-0.15$.}.  
This, combined with the fact that KELT-6 is relatively bright at
$V\sim 10.4$ (see Figure \ref{fig:rprsv}), means that this system
provides an opportunity to perform comparative measurements of two
similar planets in similar environments around stars of very different
metallicities.  In particular, we advocate attempting to acquire both
transmission and secondary eclipse spectroscopy from the ground and
space.  The resulting spectra can be compared directly with those 
already in hand for HD 209458b (e.g.,
\citealt{knutson2008,desert2008,sing2008,snellen2008,swain2009}).
Such direct comparisons may, for example, elucidate the effect of
bulk composition of the planet atmosphere on the cause of atmospheric
temperature inversions.  We note that, in order to properly plan for
secondary eclipse observations, additional radial velocity
observations will be needed to more precisely constrain the eccentricity of
KELT-6b and so predict the time of
secondary eclipse.  Such observations will also be important for
characterizing the orbit of the tertiary object in the KELT-6 system.
For these reasons, KELT-6b should prove to be a very interesting
object for further study.

\acknowledgments

K.A.C. was supported by a NASA Kentucky Space Grant Consortium Graduate Fellowship.   
Early work on KELT-North was supported by NASA Grant NNG04GO70G.
J.A.P. and K.G.S. acknowledge support from the Vanderbilt Office of the Provost through the Vanderbilt Initiative in Data-intensive Astrophysics.  
E.L.N.J. gratefully acknowledges the support of the National Science Foundation's PREST program, which helped to establish the Peter van de Kamp Observatory through grant AST-0721386.
K.G.S. acknowledges the support of the National Science Foundation through PAARE Grant AST-0849736 and AAG Grant AST-1009810.
Work by B.S.G.\ and T.G.B.\ was partially supported by NSF CAREER Grant AST-1056524.  
K.P. acknowledges support from NASA grant NNX09AB29G and NSF award number 1108686.
We thank the anonymous referee for a thoughtful reading of the manuscript and for useful suggestions. 
We thank Geoff Marcy and Howard Isaacson for Keck HIRES observations, calibration, and reduction of spectra to 1-D form.
The authors wish to recognize and acknowledge the very significant cultural role and reverence that the summit of Mauna Kea has always had within the indigenous Hawaiian community.  We are most fortunate to have the opportunity to conduct observations from this mountain. 
The TRES and KeplerCam observations were obtained with partial support from the Kepler Mission through NASA Cooperative Agreement NNX11AB99A with the Smithsonian Astrophysical Observatory (PI: D.W.L.).
A portion of this work was supported by the National Science Foundation under grant Nos. AST-1203023.
We thank Roberto Zambelli of Societa Astronomica Lunae for participation in the KELT followup network.
This work has made use of NASA's Astrophysics Data System, the Exoplanet Orbit Database at exoplanets.org, the Extrasolar Planet Encyclopedia at exoplanet.eu \citep{epe11}, the SIMBAD database operated at CDS, Strasbourg, France, and Systemic \citep{meschiari2009}.

\clearpage
\begin{deluxetable}{lcclc}
\tabletypesize{\scriptsize}
\tablecaption{KELT-6 Stellar Properties \label{tab:hostprops}}
\tablewidth{0pt}
\tablehead{\colhead{Parameter} & \colhead{Description (Units)} & \colhead{Value} & \colhead{Source} & \colhead{Ref.}}
\startdata
Names & & TYC~2532-556-1 & \\
      & & BD+31 2447 & \\
      & & Weis 32018 & \\ 
$\alpha_{\rm{J2000}}$ & &  13:03:55.647 & Tycho-2 & 1 \\
$\delta_{\rm{J2000}}$ & &  +30:38:24.26 & Tycho-2 & 1 \\
$FUV_{\rm GALEX}$ & & $20.328 \pm 0.242$ & GALEX & 2 \\
$NUV_{\rm GALEX}$ & & $14.263 \pm 0.190$ & GALEX & 2 \\
$B_T$ & & $10.837 \pm 0.049$ & Tycho-2 & 1 \\
$V_T$ & & $10.418 \pm 0.047$ & Tycho-2 & 1\\
$B_J-V_J$ & & $0.49 \pm 0.008$ & Harris & 3\\
$V$ & & $10.337 \pm 0.054$ & TASS & 4\\
$I_{\rm C}$ & & $9.745 \pm 0.061$ & TASS & 4\\
$J$ & & 9.302$\pm$0.05 & 2MASS & 5\\
$H$ & & 9.137$\pm$0.05 & 2MASS & 5\\
$K_{\rm S}$ & & 9.083$\pm$0.05 & 2MASS & 5\\
WISE1 & & 11.706$\pm$0.1 & WISE & 6\\
WISE2 & & 12.38$\pm$0.1 & WISE & 6\\
WISE3 & & 14.311$\pm$0.1 & WISE & 6\\
$\mu_{\alpha}$ & Proper Motion in RA (mas~yr$^{-1}$) \dotfill & $-6.4 \pm 0.7$ & NOMAD & 7\\
$\mu_{\delta}$ & Proper Motion in Dec. (mas~yr$^{-1}$) \dotfill & $15.6 \pm 0.7$ & NOMAD & 7\\
$\gamma_{\rm abs}$ & Absolute Systemic RV ($\kms$)\dotfill & $1.1\pm 0.2$ & This Paper\tablenotemark{a} & \\
\dotfill & Spectral Type \dotfill & F8$\pm$1 & This Paper & \\
$d$ & Distance (pc)\dotfill & $222 \pm 8$ & This Paper & \\
\dotfill & Age (Gyr)\dotfill &  $6.1 \pm 0.2$ & This Paper\tablenotemark{b} & \\
$A_V$ & Visual Extinction\dotfill & $0.01 \pm 0.02$ & This Paper & \\
($U\tablenotemark{c},V,W$) & Galactic Space Velocities (${\rm km~s}^{-1}$) \dotfill  & (-6.3$\pm$0.9, 23.2$\pm$0.8, 6.9$\pm$0.2) & This Paper\tablenotemark{d} & \\
\enddata
\tablecomments{
Magnitudes are on the AB system.
Uncertainties for the 2MASS and WISE bands were increased to 0.05 mag and 0.10 mag, respectively, to account for systematic uncertainties.
1=\citet{hog00},
2=\citet{galex05},
3=\citet{harris64},
4=\citet{tass2000},
5=\citet{skrutskie06,cutri03},
6=\citet{wright10,cutri12}.
7=\citet{zacharias04}.
}	
\tablenotetext{a}{The absolute RV uncertainty is due to the systematic uncertainties in the absolute velocities of the RV standard stars.}
\tablenotetext{b}{The uncertainty does not include possible systematic errors in the adopted evolutionary tracks.}
\tablenotetext{c}{We adopt a right-handed coordinate system such that positive $U$ is toward the Galactic Center.}
\tablenotetext{d}{See \S \ref{sec:motion}}

\end{deluxetable}

\begin{deluxetable}{lccccccc}
\tablecaption{Median Values and 68\% Confidence Intervals for Selected Physical and Orbital Parameters of the KELT-6 System from 6 Global Fits Described in \S \ref{sec:global_fits}}
\tablehead{\colhead{~~~Parameter} & \colhead{Units} & \colhead{Fit 1 (adopted)} & \colhead{Fit 2} & \colhead{Fit 3} & \colhead{Fit 4} & \colhead{Fit 5} & \colhead{Fit 6}}
\startdata
\sidehead{Global Fit Parameters:}
~~~Number of Transits\dotfill &5 or 16\dotfill & 5  & 5  & 5  &  5 & 5  & 16  \\
~~~$M_{\rm \star}$ and $R_{\rm \star}$ Constraint\dotfill &Torres or Yonsei-Yale\dotfill  & Yonsei-Yale  & Yonsei-Yale  &  Yonsei-Yale &  Yonsei-Yale & Torres  &  Yonsei-Yale \\
~~~Orbital Constraint\dotfill &Circular or Eccentric\dotfill  & Eccentric  & Circular  & Circular  & Eccentric  & Eccentric  & Eccentric  \\
~~~$\loggstar$ Prior\dotfill &Prior or No Prior\dotfill  & No Prior  & Prior  & No Prior  & Prior  & No Prior  & No Prior  \\
\sidehead{Stellar Parameters:}
~~~$\teff$\dotfill &Effective temp (K)\dotfill & $6102\pm43$   & $6101\pm43$  & $6103\pm43$  & $6102\pm44$  & $6105\pm44$  & $6109\pm44$  \\
~~~$\feh$\dotfill &Metallicity\dotfill & $-0.281_{-0.038}^{+0.039}$   & $-0.285_{-0.038}^{+0.040}$  & $-0.282_{-0.037}^{+0.039}$  & $-0.284_{-0.039}^{+0.040}$  & $-0.280_{-0.039}^{+0.039}$  & $-0.284_{-0.038}^{+0.039}$  \\
~~~$\loggstar$\dotfill &Surface gravity (cgs)\dotfill & $4.074_{-0.070}^{+0.045}$   & $4.057_{-0.037}^{+0.036}$  & $4.083_{-0.042}^{+0.022}$  & $4.012_{-0.054}^{+0.049}$  & $4.085_{-0.073}^{+0.046}$  & $4.064_{-0.068}^{+0.049}$  \\
~~~$M_{\rm \star}$\dotfill &Mass (\msun)\dotfill & $1.085_{-0.040}^{+0.043}$   & $1.086_{-0.036}^{+0.033}$  & $1.081_{-0.034}^{+0.032}$  & $1.110_{-0.041}^{+0.041}$  & $1.199_{-0.060}^{+0.066}$  & $1.090_{-0.040}^{+0.042}$  \\
~~~$R_{\rm \star}$\dotfill &Radius (\rsun)\dotfill & $1.580_{-0.094}^{+0.160}$   & $1.615_{-0.078}^{+0.086}$  & $1.562_{-0.046}^{+0.091}$  & $1.720_{-0.110}^{+0.140}$  & $1.640_{-0.100}^{+0.170}$  & $1.600_{-0.100}^{+0.160}$  \\
\sidehead{Planetary Parameters:}
~~~$M_{\rm P}$\dotfill &Mass (\mj)\dotfill & $0.430_{-0.046}^{+0.045}$   & $0.438_{-0.037}^{+0.038}$  & $0.436_{-0.037}^{+0.037}$  & $0.446_{-0.043}^{+0.044}$  & $0.461_{-0.048}^{+0.049}$  & $0.431_{-0.046}^{+0.046}$  \\
~~~$R_{\rm P}$\dotfill &Radius (\rj)\dotfill & $1.193_{-0.077}^{+0.130}$   &  $1.228_{-0.070}^{+0.080}$  &  $1.178_{-0.043}^{+0.083}$  &  $1.304_{-0.093}^{+0.110}$  &  $1.240_{-0.085}^{+0.140}$  &  $1.206_{-0.085}^{+0.120}$  \\
~~~$\loggp$\dotfill &Surface gravity\dotfill & $2.868_{-0.081}^{+0.063}$   & $2.855_{-0.061}^{+0.057}$  &  $2.885_{-0.061}^{+0.049}$ & $2.810_{-0.068}^{+0.065}$  & $2.865_{-0.083}^{+0.064}$  & $2.862_{-0.080}^{+0.064}$  \\
~~~$e$\dotfill &Eccentricity\dotfill & $0.22_{-0.10}^{+0.12}$ & $-$ & $-$ & $0.27_{-0.12}^{+0.11}$ & $0.22_{-0.10}^{+0.12}$ & $0.22_{-0.10}^{+0.12}$ \\
~~~$a$\dotfill &Semi-major axis (AU)\dotfill & $0.079_{-0.00099}^{+0.00100}$   & $0.079_{-0.00087}^{+0.00080}$   & $0.079_{-0.00085}^{+0.00078}$   & $0.080_{-0.00099}^{+0.00098}$   & $0.082_{-0.00140}^{+0.00150}$   & $0.080_{-0.001}^{+0.001}$   \\
~~~$\teq$\dotfill &Equilibrium temp (K)\dotfill & $1313_{-38}^{+59}$   & $1327_{-30}^{+33}$  & $1307_{-20}^{+34}$  & $1364_{-43}^{+48}$  & $1317_{-38}^{+61}$  &  $1323_{-41}^{+58}$ \\
\enddata
\label{tab:multipars}
\end{deluxetable}

\begin{deluxetable}{lcc}
\tablecaption{Adopted Median Values and 68\% Confidence Intervals for the Physical and Orbital Parameters of the KELT-6 System from the Fiducial Global Fit Described in \S \ref{sec:global_fits}}
\tablehead{\colhead{~~~Parameter} & \colhead{Units} & \colhead{Value (adopted)}}
\startdata
\sidehead{Stellar Parameters:}
                               ~~~$M_{\rm \star}$\dotfill &Mass (\msun)\dotfill & $1.085_{-0.040}^{+0.043}$\\
                             ~~~$R_{\rm \star}$\dotfill &Radius (\rsun)\dotfill & $1.580_{-0.094}^{+0.16}$\\
                         ~~~$L_{\rm \star}$\dotfill &Luminosity (\lsun)\dotfill & $3.11_{-0.39}^{+0.68}$\\
                             ~~~$\rho_{\rm \star}$\dotfill &Density (cgs)\dotfill & $0.387_{-0.088}^{+0.068}$\\
                  ~~~$\loggstar$\dotfill &Surface gravity (cgs)\dotfill & $4.074_{-0.070}^{+0.045}$\\
                  ~~~$\teff$\dotfill &Effective temperature (K)\dotfill & $6102\pm43$\\
                                 ~~~$\feh$\dotfill &Metallicity\dotfill & $-0.281_{-0.038}^{+0.039}$\\
\sidehead{Planetary Parameters:}
                                   ~~~$e$\dotfill &Eccentricity\dotfill & $0.22_{-0.10}^{+0.12}$\\
        ~~~$\omega_{\rm \star}$\dotfill &Argument of periastron (degrees)\dotfill & $80_{-120}^{+110}$\\
                                  ~~~$P$\dotfill &Period (days)\dotfill & $7.8457\pm0.0002$\\
                           ~~~$a$\dotfill &Semi-major axis (AU)\dotfill & $0.07939_{-0.00099}^{+0.0010}$\\
                                 ~~~$M_{\rm P}$\dotfill &Mass (\mj)\dotfill & $0.430_{-0.046}^{+0.045}$\\
                               ~~~$R_{\rm P}$\dotfill &Radius (\rj)\dotfill & $1.193_{-0.077}^{+0.13}$\\
                           ~~~$\rho_{\rm P}$\dotfill &Density (cgs)\dotfill & $0.311_{-0.076}^{+0.069}$\\
                      ~~~$\loggp$\dotfill &Surface gravity\dotfill & $2.868_{-0.081}^{+0.063}$\\
               ~~~$\teq$\dotfill &Equilibrium temperature (K)\dotfill & $1313_{-38}^{+59}$\\
                           ~~~$\Theta$\dotfill &Safronov number\dotfill & $0.0521_{-0.0061}^{+0.0059}$\\
                   ~~~$\fave$\dotfill &Incident flux (\fluxcgs)\dotfill & $0.653_{-0.076}^{+0.092}$\\
\sidehead{RV Parameters:}
       ~~~$T_{\rm C}$\dotfill &Time of inferior conjunction (\bjdtdb)\dotfill & $2456269.3399_{-0.0072}^{+0.0071}$\\
               ~~~$T_{\rm P}$\dotfill &Time of periastron (\bjdtdb)\dotfill & $2456269.2_{-2.5}^{+1.7}$\\
                        ~~~$K$\dotfill &RV semi-amplitude (m s$^{-1}$)\dotfill & $42.8_{-4.2}^{+4.5}$\\
                    ~~~$M_{\rm P}\sin{i}$\dotfill &Minimum mass (\mj)\dotfill & $0.430_{-0.046}^{+0.045}$\\
                           ~~~$M_{\rm P}/M_{\rm \star}$\dotfill &Mass ratio\dotfill & $0.000378_{-0.000037}^{+0.000036}$\\
                       ~~~$u$\dotfill &RM linear limb darkening\dotfill & $0.6035_{-0.0039}^{+0.0040}$\\
                               ~~~$\gamma_{\rm HIRES}$\dotfill &m s$^{-1}$\dotfill & $-3.1\pm3.2$\\
                  ~~~$\dot{\gamma}_{\rm HIRES}$\dotfill &RV slope (m s$^{-1}$ day$^{-1}$)\dotfill & $-0.239\pm0.037$\\
                                         ~~~$\ecosw$\dotfill & \dotfill & $0.02_{-0.14}^{+0.13}$\\
                                         ~~~$\esinw$\dotfill & \dotfill & $0.05_{-0.22}^{+0.23}$\\
                     ~~~$f(m1,m2)$\dotfill &Mass function (\mj)\dotfill & $0.000000061_{-0.000000017}^{+0.000000020}$\\
\sidehead{Primary Transit Parameters:}
~~~$R_{\rm P}/R_{\rm \star}$\dotfill &Radius of the planet in stellar radii\dotfill & $0.07761_{-0.00092}^{+0.0010}$\\
           ~~~$a/R_{\rm \star}$\dotfill &Semi-major axis in stellar radii\dotfill & $10.79_{-0.89}^{+0.60}$\\
                          ~~~$i$\dotfill &Inclination (degrees)\dotfill & $88.81_{-0.91}^{+0.79}$\\
                               ~~~$b$\dotfill &Impact parameter\dotfill & $0.20_{-0.13}^{+0.14}$\\
                             ~~~$\delta$\dotfill &Transit depth\dotfill & $0.00602_{-0.00014}^{+0.00016}$\\
                      ~~~$T_{\rm 0}$\dotfill &Best-fit linear ephemeris from transits (\bjdtdb)\dotfill & $2456347.796793 \pm 0.000364$\\
                      ~~~$P_{\rm Transit}$\dotfill &Best-fit linear ephemeris period from transits (days)\dotfill & $7.8456314 \pm 0.0000459$\\
                    ~~~$T_{\rm FWHM}$\dotfill &FWHM duration (days)\dotfill & $0.212_{-0.029}^{+0.039}$\\
              ~~~$\tau$\dotfill &Ingress/egress duration (days)\dotfill & $0.0175_{-0.0028}^{+0.0039}$\\
                     ~~~$T_{\rm 14}$\dotfill &Total duration (days)\dotfill & $0.230_{-0.032}^{+0.043}$\\
   ~~~$P_{\rm T}$\dotfill &A priori non-grazing transit probability\dotfill & $0.091_{-0.021}^{+0.038}$\\
              ~~~$P_{\rm T,G}$\dotfill &A priori transit probablity\dotfill & $0.107_{-0.024}^{+0.044}$\\
\sidehead{Secondary Eclipse Parameters:}
                  ~~~$T_{\rm S}$\dotfill &Time of eclipse (\bjdtdb)\dotfill & $2456265.51_{-0.70}^{+0.66}$\\
                           ~~~$b_{\rm S}$\dotfill &Impact parameter\dotfill & $0.22_{-0.14}^{+0.18}$\\
                  ~~~$T_{\rm S,FWHM}$\dotfill &FWHM duration (days)\dotfill & $0.231_{-0.051}^{+0.073}$\\
            ~~~$\tau_S$\dotfill &Ingress/egress duration (days)\dotfill & $0.0194_{-0.0048}^{+0.0083}$\\
                   ~~~$T_{\rm S,14}$\dotfill &Total duration (days)\dotfill & $0.251_{-0.056}^{+0.081}$\\
   ~~~$P_{\rm S}$\dotfill &A priori non-grazing eclipse probability\dotfill & $0.084_{-0.010}^{+0.018}$\\
             ~~~$P_{\rm S,G}$\dotfill &A priori eclipse probability\dotfill & $0.098_{-0.012}^{+0.020}$
\enddata
\label{tab:fitpars}
\end{deluxetable}


\begin{thebibliography}{}

\bibitem[Alard(2000)]{al00}
Alard, C.\ 2000, \aaps, 144, 363

\bibitem[Alard \& Lupton(1998)]{al98}
Alard, C., \& Lupton, R. H.\ 1998, \apj, 503, 325

\bibitem[Alonso et al.(2004)]{alon04}
Alonso, R., Brown, T.~M., Torres, G., et al.\ 2004, \apjl, 613, L153     

\bibitem[Alsubai et al.(2011)]{al11}
Alsubai, K.~A., Parley, N.~R., Bramich, D.~M., et al.\ 2011, \mnras, 417, 709 

\bibitem[Anderson et al.(2013)]{anderson2013} Anderson, D.~R., Collier Cameron, A., Delrez, L., et al.\ 2013, arXiv:1310.5654

\bibitem[Assef et al.(2009)]{assef09}
Assef, R.~J., Gaudi, B.~S., \& Stanek, K.~Z.\ 2009, \apj, 701, 1616

\bibitem[Baglin(2003)]{bag03}
Baglin, A.\ 2003, Advances in Space Research, 31, 345 

\bibitem[Bakos et al.(2007)]{bakos07} 
Bakos, G.~{\'A}., Noyes, R.~W., Kov{\'a}cs, G., et al.\ 2007, \apj, 656, 552 

\bibitem[Baraffe et al.(1998)]{baraffe1998}
Baraffe, I., Chabrier, G., Allard, F., \& Hauschildt, P.~H.\ 2008, \aap, 337, 403

\bibitem[Barnes(2007)]{barnes07}
Barnes, S.~A.\ 2007, \apj, 669, 1167

\bibitem[Beatty et al.(2012)]{beatty12} Beatty, T.~G., Pepper, J., Siverd, R.~J., et al.\ 2012, \apjl, 756, L39

\bibitem[Borucki et al.(2010)]{boru10}
Borucki, W.~J., Koch, D., Basri, G., et al.\ 2010, Science, 327, 977 

\bibitem[Brown(2003)]{brown2003} Brown, T.~M.\ 2003, \apjl, 593, L125

\bibitem[Brown et al.(2013)]{brown13} Brown, T.~M., Baliber, N., Bianco, F., et al.\ 2013, \pasp, 125, 1031

\bibitem[Buchhave et al.(2010)]{bu10}
Buchhave, L.~A., Bakos, G.~{\'A}., Hartman, J.~D., et al.\ 2010, \apj, 720, 1118

\bibitem[Buchhave et al.(2012)]{bu12}
Buchhave, L.~A., Latham, D.~W., Johansen, A., et al.\ 2012, \nat, 486, 375

\bibitem[Burrows et al.(2007)]{burrows2007}
Burrows, A., Hubeny, I., Budaj, J., \& Hubbard, W.~B.\ 2007, \apj, 661, 502

\bibitem[Butler et al.(1996)]{butler96}
Butler, R.~P., Marcy, G.~W., Williams, E., McCarthy, C., Dosanjh, P., Vogt, S.~S.\ 1996, \pasp, 108, 500

\bibitem[Charbonneau et al.(2000)]{charbonneau00} Charbonneau, D., 
Brown, T.~M., Latham, D.~W., \& Mayor, M.\ 2000, \apjl, 529, L45 

\bibitem[Chubak et al.(2012)]{chubak2012}
Chubak, C., Marcy, G.~W., Fischer, D.~A., et al.\ 2012, arXiv:1207.6212

\bibitem[Claret \& Bloemen(2011)]{cb11}
Claret, A., \& Bloemen, S.\ 2011, \aap, 529, A75

\bibitem[Collier Cameron et al.(2007a)]{cc07a}
Collier Cameron, A., Bouchy, F., H{\'e}brard, G., et al.\ 2007, \mnras, 375, 951 

\bibitem[Collier Cameron et al.(2007b)]{cc07b}
Collier Cameron, A., Wilson, D.~M., West, R.~G., et al.\ 2007, \mnras, 380, 1230

\bibitem[Crepp et al.(2012)]{crepp2012}
Crepp, J.~R., Johnson, J.~A., Howard, A.~W., et al.\ 2012, \apj, 761, 39

\bibitem[Cutri et al.(2003)]{cutri03} Cutri, R.~M., Skrutskie,
M.~F., van Dyk, S., et al.\ 2003, yCat, 2246, 0

\bibitem[Cutri et al.(2012)]{cutri12} Cutri, R.~M., Wright, E.~L., Bauer, J., et al.\ 2012, yCat, 2311, 0

\bibitem[Dawson \& Murray-Clay(2013)]{dawson2013} Dawson, R.~I., \& Murray-Clay, R.~A.\ 2013, \apj, 767, 24

\bibitem[Demarque et al.(2004)]{yy04}
Demarque, P., Woo, J.-H., Kim, Y.-C., \& Yi, S.~K.\ 2004, \apjs, 155, 667

\bibitem[Demory \& Seager(2011)]{ds11} Demory, B.-O., \& Seager, S.\ 2011, \apjs, 197, 12

\bibitem[D{\'e}sert et 
al.(2008)]{desert2008} D{\'e}sert, J.-M., Vidal-Madjar, A., Lecavelier Des Etangs, A., et al.\ 2008, \aap, 492, 585 

\bibitem[Eastman et al.(2010)]{east10} 
Eastman, J., Siverd, R., \& Gaudi, B.~S.\ 2010, \pasp, 122, 935 

\bibitem[Eastman et al.(2013)]{eastman13} 
Eastman, J., Gaudi, B.~S., \& Agol, E.\ 2013, \pasp, 125, 83 

\bibitem[ESA(1997)]{esa97} ESA. 1997, The Hipparcos and Tycho Catalogues (ESA SP-1200; Noordwijk: ESA)

\bibitem[Ford(2006)]{ford2006}
Ford, E.B.\ 2006, ApJ, 642, 505

\bibitem[Fulton et al.(2011)]{fulton2011}
Fulton, B.~J., Shporer, A., Winn, J.~N., Holman, M.~J.,  P{\'a}l, A., \& Gazak, J.~Z.\ 2011, \aj, 142, 84

\bibitem[F{\H u}r{\'e}sz(2008)]{tres}
F{\H u}r{\'e}sz, G.\ 2008, PhD thesis, Univ. Szeged

\bibitem[Gould \& Morgan(2003)]{gm03} 
Gould, A., \& Morgan, C.~W.\ 2003, \apj, 585, 1056 

\bibitem[Harris \& Upgren(1964)]{harris64}
Harris, D.~L., \& Upgren, A.~R.\ 1964, \apj, 140, 151

\bibitem[Hartman et al.(2008)]{hart08}
Hartman, J.~D., Gaudi, B.~S., Holman, M.~J., et al.\ 2008, \apj, 675, 1254

\bibitem[Hauschildt et al.(1999)]{hau99}
Hauschildt, P.~H., Allard, F., Ferguson, J., Baron, E., \& Alexander, D.~R.\ 1999, \apj, 525, 871

\bibitem[Henry et al.(2000)]{henry00}
Henry, G.~W., Marcy, G.~W., Butler, R.~P., Baron, E., \& Vogt, S.~S.\ 2000, \apjl, 529, L41

\bibitem[ESA(1997)]{esa97} 
The Hipparcos and Tycho Catalogues\ 1997,  ESA SP-1200  

\bibitem[H{\o}g et al.(2000)]{hog00}
H{\o}g, E., Fabricius, C., Makarov, V.~V., et al.\ 2000, \aap, 355, L27

\bibitem[Howard et al.(2011)]{howard11}
Howard, A.~W., Johnson, J.~A., Marcy, G.~W., et al.\ 2011, \apj, 726, 73

\bibitem[Jensen(2013)]{jensen2013}
Jensen, E. L. N. 2013, Tapir, Astrophysics Source Code Library, record ascl:1306.007

\bibitem[Johnson et al.(2010)]{johnson10}
Johnson, J.~A., Howard, A.~W., Marcy, G.~W., et al.\ 2010, \pasp, 122, 149

\bibitem[Knutson et al.(2008)]{knutson2008} Knutson, H.~A., 
Charbonneau, D., Allen, L.~E., Burrows, A., 
\& Megeath, S.~T.\ 2008, \apj, 673, 526 

\bibitem[Kov{\'a}cs et al.(2002)]{k02} Kov{\'a}cs, G., Zucker, S., \& Mazeh, T.\ 2002, \aap, 391, 369

\bibitem[Kov{\'a}cs et al.(2005)]{k05}
Kov{\'a}cs, G., Bakos, G., \& Noyes, R.~W.\ 2005, \mnras, 356, 557


\bibitem[Latham et al.(2009)]{latham2009} Latham, D.~W., Bakos, G.~A., Torres, G., et al.\ 2009, \apj, 704, 1107

\bibitem[Lucy \& Sweeney(1971)]{lsbias}
Lucy, L.~B., \& Sweeney, M.~A.\ 1971, \aj, 76, 544

\bibitem[Madhusudhan 
\& Seager(2010)]{madhu2010} Madhusudhan, N., \& Seager, S.\ 2010, \apj, 725, 261 

\bibitem[Marcy  \& Butler(1992)]{mb92} 
Marcy, G.~W., \& Butler, R.~P.\ 1992, \pasp 104, 270 

\bibitem[Martin et al.(2005)]{galex05}
Martin, D.~C., Fanson, J., Schiminovich, D., et al.\ 2005, \apjl, 619, L1

\bibitem[McCullough et al.(2006)]{mc06}
McCullough, P.~R., Stys, J.~E., Valenti, J.~A., et al.\ 2006, \apj, 648, 1228 

\bibitem[Meschiari et al.(2009)]{meschiari2009}
Meschiari, S., Wolf, A.~S., Rivera, E., et al.\ 2009, \pasp, 121, 1016 

\bibitem[O'Donovan et al.(2006)]{odonovan2006} O'Donovan, F.~T., 
Charbonneau, D., Torres, G., et al.\ 2006, \apj, 644, 1237 

\bibitem[Pepper et al.(2003)]{pepper03} Pepper, J., Gould, A., \& DePoy, D.~L.\ 2003, Acta Astronomica, 53, 213

\bibitem[Pepper et al.(2007)]{pepper07} Pepper, J., Pogge, R.~W., DePoy, D.~L., et al.\ 2007, \pasp, 119, 923

\bibitem[Pepper et al.(2013)]{pepper13} Pepper, J., Siverd, R.~J., Beatty, T.~G., et al.\ 2013, \apj, 773, 64

\bibitem[Ribas 
\& Miralda-Escud{\'e}(2007)]{ribas2007} Ribas, I., \& Miralda-Escud{\'e}, J.\ 2007, \aap, 464, 779 

\bibitem[Richmond et al.(2000)]{tass2000}
Richmond, M.~W., Droege, T.~F., Gombert, G., et al.\ 2000, \pasp, 112, 397

\bibitem[Sato et al.(2005)]{sato2005}
Sato, B., Fischer, D.~A., Henry, G.~W., Laughlin, G., et al.\ 2005, \apj, 633, 465

\bibitem[Schneider et  al.(2011)]{epe11}
Schneider, J., Dedieu, C., Le Sidaner, P., Savalle, R., \& Zolotukhin, I.\ 2011, \aap, 532, A79

\bibitem[Siess et al.(2000)]{siess00}
Siess, L., Dufour, E., \& Forestini, M.\ 2000, \aap, 358, 593

\bibitem[Sing et al.(2008)]{sing2008} Sing, D.~K., Vidal-Madjar, 
A., D{\'e}sert, J.-M., Lecavelier des Etangs, A., 
\& Ballester, G.\ 2008, \apj, 686, 658 

\bibitem[Siverd et al.(2012)]{siv12}
Siverd, R.~J., Beatty, T.~G., Pepper, J., et al.\ 2012, \apj, 761, 123 

\bibitem[Skrutskie et al.(2006)]{skrutskie06} Skrutskie, M.~F.,
Cutri, R.~M., Stiening, R., et al.\ 2006, \aj, 131, 1163

\bibitem[Snellen et 
al.(2008)]{snellen2008} Snellen, I.~A.~G., Albrecht, S., de Mooij, E.~J.~W., \& Le Poole, R.~S.\ 2008, \aap, 487, 357 

\bibitem[Spiegel \& Madhusudhan(2012)]{sm12} Spiegel, D.~S., \& Madhusudhan, N.\ 2012, \apj, 756, 132

\bibitem[Stetson(1987)]{stet87}
Stetson, P.~B.\ 1987, \pasp, 99, 191

\bibitem[Swain et al.(2009)]{swain2009} Swain, M.~R., Tinetti, 
G., Vasisht, G., et al.\ 2009, \apj, 704, 1616 

\bibitem[Torres(1999)]{torres99} Torres, G.\ 1999, \pasp, 111, 169

\bibitem[Torres et al.(2007)]{torres2007} Torres, G., Bakos, G.~{\'A}., Kov{\'a}cs, G., et al.\ 2007, \apjl, 666, 121

\bibitem[Torres et al.(2008)]{torres2008} Torres, G., Winn, J.~N., \& Holman, M.~J.\ 2008, \apj, 677, 1324

\bibitem[Torres et al.(2010)]{torres10} Torres, G., Andersen, J., \& Gim{\'e}nez, A.\ 2010, \aapr, 18, 67

\bibitem[Valenti \& Piskunov(1996)]{vp96}
Valenti, J.~A., \& Piskunov, N.\ 1996, \aaps, 118, 595

\bibitem[Valenti \& Fischer(2005)]{vf05}
Valenti, J.~A., \& Fischer, D.~A.\ 2005, \apjs, 159, 141

\bibitem[Vogt et al.(1994)]{vogt94} Vogt, S.~S., Allen, S.~L., Bigelow, B.~C., et al.\ 1994, Proc. SPIE, 2198, 362

\bibitem[Winn(2009)]{winn2009} Winn, J.~N.\ 2009, in IAU Symp. 253, Transiting Planets, ed. F. Pont, D. D. Sasselov, \& M. J. Holman (Cambridge: Cambridge Univ. Press), 99

\bibitem[Winn(2010)]{winn2010} 
Winn, J.~N.\ 2010, in Exoplanets, ed. S. Seager, Tucson:University of Arizona Press, 55 

\bibitem[Wright et al.(2010)]{wright10} Wright, E.~L.,
Eisenhardt, P.~R.~M., Mainzer, A.~K., et al.\ 2010, \aj, 140, 1868

\bibitem[Wright et al.(2011)]{wright11} Wright, E.~L.,
 Fakhouri, O., Marcy, G.~W., et al.\ 2011, \pasp, 123, 412

\bibitem[Wright(2005)]{wright05} Wright, J.~T\ 2012, \pasp, 117, 657 

\bibitem[Zacharias et al.(2004)]{zacharias04} Zacharias, N., Monet,
D.~G., Levine, S.~E., et al.\ 2004, Bulletin of the American Astronomical
Society, 36, 1418

\end{thebibliography}
\end{document}